\documentclass[12pt]{iopart}

\pdfoutput=1

\usepackage{graphicx,sidecap}
\usepackage{bm}
\usepackage{braket}
\usepackage{amssymb}
\usepackage{mathrsfs}
\usepackage{amsmath}
\usepackage{xcolor}
\usepackage{hyperref}
\usepackage{enumerate}
\usepackage[toc,title]{appendix}

\hypersetup{colorlinks,bookmarksopen,bookmarksnumbered,
citecolor=red,
linkcolor=blue,
pdfstartview=green,
urlcolor=purple}

\catcode`@=11 
\renewcommand\tableofcontents{%
  \section*{\contentsname}%
  \@starttoc{toc}%
}
\catcode`@=12

\def\be{\begin{equation}}
\def\ee{\end{equation}}

\def\bea{\begin{eqnarray}}
\def\eea{\end{eqnarray}}

\def\Tr{{\rm Tr}}

\usepackage{xparse}

\DeclareDocumentCommand{\TrProd}{ m O{} o O{} o O{} o O{} o }{%
\{ \Gamma_{#1}^{#2}
\IfNoValueTF{#3}{}{,\Gamma_{#3}^{#4}}
\IfNoValueTF{#5}{}{,\Gamma_{#5}^{#6}}
\IfNoValueTF{#7}{}{,\Gamma_{#7}^{#8}}
\IfNoValueTF{#9}{}{,\Gamma_{#9}}
\}
}

\DeclareDocumentCommand{\TrProdTilde}{ m O{} o O{} o O{} o O{} o }{%
\{ \widetilde\Gamma_{#1}^{#2}
\IfNoValueTF{#3}{}{,\widetilde\Gamma_{#3}^{#4}}
\IfNoValueTF{#5}{}{,\widetilde\Gamma_{#5}^{#6}}
\IfNoValueTF{#7}{}{,\widetilde\Gamma_{#7}^{#8}}
\IfNoValueTF{#9}{}{,\widetilde\Gamma_{#9}}
\}
}

\DeclareDocumentCommand{\ch}{ m o o o m o o o }{%
\begin{bmatrix}
#1 \IfNoValueTF{#2}{}{& #2}\IfNoValueTF{#3}{}{& #3}\IfNoValueTF{#4}{}{& #4} \\
#5 \IfNoValueTF{#6}{}{& #6}\IfNoValueTF{#7}{}{& #7}\IfNoValueTF{#8}{}{& #8}
\end{bmatrix}_{\tau}
}

\makeatletter
\newcommand{\pushright}[1]{\ifmeasuring@#1\else\omit\hfill$\displaystyle#1$\fi\ignorespaces}
\newcommand{\pushleft}[1]{\ifmeasuring@#1\else\omit$\displaystyle#1$\hfill\fi\ignorespaces}
\makeatother

\begin{document}

\title[
Entanglement negativity in a two dimensional harmonic lattice
]{
\\
Entanglement negativity in a two dimensional harmonic lattice:
Area law and corner contributions
}

\vspace{.5cm}

\author{Cristiano De Nobili$^1$, Andrea Coser$^2$ and Erik Tonni$^1$}
\address{$^1$\,SISSA and INFN, via Bonomea 265, 34136 Trieste, Italy. }
\address{$^2$\,Departamento de An\'alisis Matem\'atico, Universidad Complutense de Madrid, 28040 Madrid, Spain.}

\vspace{.5cm}

\begin{abstract}
We study the logarithmic negativity and the moments of the partial transpose in the ground state of a two dimensional massless harmonic square lattice with nearest neighbour interactions for various configurations of adjacent  domains.
At leading order for large domains, the logarithmic negativity and the logarithm of the ratio between the generic moment of the partial transpose and the moment of the reduced density matrix at the same order satisfy an area law in terms of the length of the curve shared by the adjacent regions. 
We give numerical evidences that the coefficient of the area law term in these quantities is related to the coefficient of the area law term in the R\'enyi entropies.
Whenever the curve shared by the adjacent domains contains vertices, a subleading logarithmic term occurs in these quantities and the numerical values of the corner function for some pairs of angles are obtained.
In the special case of vertices corresponding to explementary angles, we provide numerical evidence that the corner function of the logarithmic negativity is given by the corner function of the  R\'enyi entropy of order 1/2.
\end{abstract}

\maketitle

\tableofcontents

\section{Introduction}
\label{sec:intro}

The entanglement in extended quantum systems and the ways to quantify it has attracted a lot of research during the last decade in different areas of theoretical physics like condensed matter, quantum information and quantum gravity (see \cite{rev} for reviews).

Given a quantum system in a pure state $| \Psi \rangle$ and assuming that its Hilbert space is bipartite, i.e. $\mathcal{H} = \mathcal{H}_A \otimes \mathcal{H}_B$, 
the entanglement entropy is an important quantity to measure the entanglement between the two parts of the system. 
In this manuscript we consider only spatial bipartitions, denoting by $A$ a generic spatial region and by $B$ its complement. 
The  $A$'s reduced density matrix $\rho_A$ is obtained by tracing over $ \mathcal{H}_B$ the density matrix $\rho= | \Psi \rangle \langle \Psi |$ of the entire system. 
The normalisation condition is  $\Tr \rho_A =1$ and the Von Neumann entropy of $\rho_A$ defines the entanglement entropy, namely  
\be
\label{SA def}
S_A = -\, \Tr (\rho_A \log \rho_A)\,.
\ee
 The entanglement entropy  $S_B$ obtained from $B$'s reduced density matrix can be introduced in the same way.
 When $\rho$ is a pure state, we have that $S_B = S_A$.

Important quantities to study the bipartite entanglement of pure states are also the R\'enyi entropies, which are defined as follows
\be
\label{Renyi entropies def}
S_A^{(n)} = \frac{1}{1-n} \, \log \Tr \rho_A^n \,,
\ee
where $n\geqslant 2$ is an integer parameter.
The entanglement entropy (\ref{SA def}) can be found from the R\'enyi entropies (\ref{Renyi entropies def}) as $S_A = - \lim_{n \rightarrow 1} \partial_n \Tr \rho_A^n = \lim_{n \rightarrow 1} S_A^{(n)}$. This limit requires to perform an analytical continuation of  $S_A^{(n)}$ to real values of $n$ close to $n=1$.
For a pure state $S^{(n)}_B = S^{(n)}_A$ for any value of $n$.

When the subsystem $A = A_1 \cup A_2$ is made by two regions, which can be either disjoint or adjacent, an interesting quantity to consider is the mutual information, i.e.
\be
\label{MI def}
I_{A_1, A_2} = S_{A_1} + S_{A_2} - S_{A_1 \cup A_2} 
=  \lim_{n \to 1} I_{A_1,A_2}^{(n)}\,,
\ee
where  $I_{A_1, A_2}^{(n)} = S_{A_1}^{(n)} + S_{A_2}^{(n)} - S_{A_1 \cup A_2}^{(n)} $ is the corresponding combination of R\'enyi entropies.
The mutual information measures the total amount of correlations between the two systems \cite{MI properties, wvhc-08-proof area law}.

For some quantum systems on the lattice, the entanglement entropy $S_A$ grows like the area of the boundary of the subsystem $A$ as its size increases \cite{area law 0, area law 1}.
This area law behaviour of $S_A$ has been proved for gapped systems on the lattice \cite{wvhc-08-proof area law}, but for critical systems the situation is more complicated: important exceptions are the critical systems in one spatial dimension, where $S_A$ diverges logarithmically with the length of the interval $A$ \cite{vlrk-03}, and free fermions in higher dimensions \cite{area law violation fermions}.

When the continuum limit is described by a quantum field theory, an ultraviolet (UV) cutoff $\varepsilon$ must be introduced and $S_A$ is a divergent quantity for $\varepsilon  \to 0$.
The area law behaviour can occur in the coefficient of the most divergent term, which is non universal and turns out to be proportional to the area of $\partial A$ (i.e. the boundary of $A$) in these cases. 
A quantum system in the continuum at criticality is described by a conformal field theory (CFT).
Considering a $1+1$ dimensional CFT on a line at zero temperature and an interval $A$ of length $\ell$, we have that $S_A = (c/3) \log(\ell/\varepsilon) + \textrm{const}$, where $c$ is the central charge of the model \cite{ee cft 1 interval}.
Instead, the mutual information (\ref{MI def}) of two disjoint intervals is UV finite and it depends on the full operator content of the model \cite{2 disjoint intervals}.
Performing the replica limit to get analytic expressions for $S_A$ and $I_{A_1, A_2} $ from the ones for $S_A^{(n)}$ and $I^{(n)}_{A_1, A_2} $ can be a very difficult task (see \cite{rational interpolation, rational interpolation-noi} for a numerical approach).

In $2+1$ dimensional CFTs and for domains $A$ whose boundary is smooth, the expansion of the R\'enyi entropies reads $S_A^{(n)} = \tilde{\alpha}_n \, P_A/\varepsilon + \textrm{const}$ as $\varepsilon \to 0$, where $P_A$ is the perimeter of $A$ and the coefficient $\tilde{\alpha}_n$ depends on the model and on the details of the UV regularisation.
The replica limit implies that $S_A = \tilde{\alpha} \,P_A/\varepsilon + \textrm{const}$ as $\varepsilon \to 0$, where $\lim_{n\to 1} \tilde{\alpha}_n  = \tilde{\alpha}$.
When the two dimensional spatial domain $A$ has a non smooth boundary, $S^{(n)}_A$ contains also a subleading logarithmic term whose coefficient is independent of the regularisation details.
Such coefficient is obtained as the sum of the contributions of the vertices of the curve $\partial A$, where each term is given by a model dependent function (corner function) evaluated on the opening angle in $A$ of the corresponding vertex. 
Many interesting studies have been done on such corner functions in various lattice models \cite{ch-06-corners, chl-08-corners, ch-review, corner-lattice}.
For a CFT in $2+1$ dimensions, it has been recently found that the leading term in the expansion of the corner function as the opening angle $\theta$ is close to $\pi$ provides the constant characterising the two-point function of the stress tensor \cite{corner-recent qft,bmk-jhep}.
Within the context of the AdS/CFT correspondence, the prescription to compute holographically the  
entanglement entropy in the regime where classical gravity can be employed has been found in \cite{RT}. 
In the case of AdS$_4$/CFT$_3$, such holographic prescription provides also the expected subleading logarithmic divergence whenever $\partial A$ contains vertices \cite{holography-corner}.

The mutual information (\ref{MI def}) of disjoint domains with smooth boundaries is a UV finite quantity because the area law terms cancel.
When the separation between the regions is large with respect to their sizes, analytic results have been found \cite{cardy-some results}.
In the case of two dimensional adjacent domains $A_1$ and $A_2$, both the mutual information and its generalisation involving the R\'enyi entropies display an area law behaviour in terms of 
the length of the curve shared by the adjacent regions. In particular, $I_{A_1, A_2}^{(n)} = 2\tilde{\alpha}_n P_{\textrm{\tiny shared}}/\varepsilon + \dots$ and $I_{A_1, A_2} = 2\tilde{\alpha}\, P_{\textrm{\tiny shared}}/\varepsilon + \dots$  as $\varepsilon \to 0$, where $P_{\textrm{\tiny shared}} \equiv \textrm{length}(\partial A_1 \cap \partial A_2)$.

The entanglement entropy and the R\'enyi entropies are measures of the quantum entanglement for a bipartite system in a pure state, 
but this is not true when the whole system is in a mixed state.
For instance, the mutual information of a bipartite system in a thermal state is dominated by classical correlations.
Another important example of mixed state is the reduced density matrix $\rho_A$ associated to a subsystem $A$, when the entire system is in its ground state. 
Splitting $A$ in two domains $A_1$ and $A_2$, which can be either adjacent or disjoint, it is worth considering the bipartite entanglement between them.
Many measures of the bipartite entanglement for a mixed state have been proposed in quantum information theory, but they are usually very difficult to compute, even for small systems. A measure which is computable also for extended systems is  the logarithmic negativity \cite{neg papers}.

Let us consider a mixed state characterised by the density matrix $\rho$ acting on a spatially bipartite Hilbert space $\mathcal{H} = \mathcal{H}_{A_1} \otimes \mathcal{H}_{A_2}$. 
We remark that  $\mathcal{H}$ can be either the Hilbert space characterising the whole spatial system or the one $\mathcal{H}_A$ associated to the bipartite subsystem $A = A_1\cup A_2$ introduced above (in the latter case $\rho=\rho_A$).
The logarithmic negativity is defined through the partial transpose of $\rho$ with respect to one of the two parts.
Considering e.g. the partial transposition with respect to $A_2$, the matrix element of $\rho^{T_2}$ is given by
\be
\label{PartialTransp}
\langle e_i^{(1)} e_j^{(2)}| \rho^{T_2}| e_k^{(1)} e_l^{(2)}\rangle = \langle e_i^{(1)} e_l^{(2)}| \,\rho\,| e_k^{(1)} e_j^{(2)}\rangle\,.
\ee
Since the spectrum of the Hermitian matrix $\rho^{T_2}$ can contain also negative eigenvalues, it is worth computing its trace norm $ \| \rho_A^{T_2} \| = \Tr |\rho^{T_2}| = \sum_i |\lambda_i|$.
The logarithmic negativity is defined as 
\be
\label{logneg def}
\mathcal{E}
\equiv
\log \Tr |\rho^{T_2}|\,.
\ee

The logarithmic negativity can be computed also by employing a replica limit \cite{cct-neg-letter,cct-neg-long}.
Considering the $n$-th moment $\Tr (\rho^{T_2} )^n$ of the partial transpose and taking into account only the sequence of the even powers $n=n_e$, it is not difficult to realise that (\ref{logneg def}) can be found by performing the following analytic continuation 
\be
\label{replica limit neg}
\mathcal{E} = \lim_{n_e \to 1} 
\log \Tr \big(\rho^{T_2} \big)^{n_e} .
\ee

In the special case of a pure state $\rho = |\Psi \rangle \langle \Psi |$ acting on a bipartite Hilbert space, the moments of the partial transpose are related to the R\'enyi entropies as follows \cite{cct-neg-letter,cct-neg-long}
\be
\label{identity pure states}
\Tr \big(|\Psi \rangle \langle \Psi |^{T_2} \big)^{n} 
\,=\,
\Bigg\{ \begin{array}{ll}
\Tr \rho_{A_2}^{n_o}  \hspace{1.5cm} &  \textrm{odd $n = n_o$}\,,
\\
\rule{0pt}{.6cm}
\big(\Tr \rho_{A_2}^{n_e/2} \big)^2 &  \textrm{even $n = n_e$}\,,
\end{array}
\ee
where $\rho_{A_2} = \Tr_{A_1} |\Psi \rangle \langle \Psi |$ is the reduced density matrix of the subsystem $A_2$.
From the relation (\ref{identity pure states}) and the replica limit (\ref{replica limit neg}), one easily gets that $\mathcal{E} = S^{(1/2)}_{A_2}$.

The logarithmic negativity and the moments $\Tr (\rho_A^{T_2})^n$ of the partial transpose are interesting quantities to compute for bipartite mixed states.
In this manuscript we focus on a particular system in its ground state, considering the mixed state given by the reduced density matrix $\rho_A = \Tr_{\mathcal{H}_B} | \Psi \rangle \langle \Psi |$ of a spatial subsystem, whose corresponding Hilbert space $\mathcal{H}_A = \mathcal{H}_{A_1} \otimes \mathcal{H}_{A_2}$ is bipartite.
Instead of the moments of $\rho_A^{T_2}$, we find more convenient to consider the following quantity
\be
\label{En def}
\mathcal{E}_n \,\equiv\, \log \left( \frac{\Tr (\rho_A^{T_2})^n}{\Tr \rho_A^n} \right) .
\ee
It is not difficult to show that $\mathcal{E}_2=0$.
Given the normalization condition of $\rho_A$, the replica limit (\ref{replica limit neg}) tells that
\be
\label{replica limit E_n}
\mathcal{E} = \lim_{n_e \to 1} \mathcal{E}_{n_e}\,.
\ee

For $1+1$ dimensional CFTs in the ground state, the logarithmic negativity and the moments of the partial transpose have been studied for both adjacent and disjoint intervals \cite{cct-neg-letter, cct-neg-long, neg-after}. 
This analysis has been extended also to a bipartite system at finite temperature \cite{cct-neg-T}.
The moments of the partial transpose for some fermionic systems on the lattice have been studied through a method involving correlators in \cite{ez-15, ctc-15} and the overlap matrix in \cite{cw-16}.
The logarithmic negativity has been considered also for a non vanishing mass \cite{cct-neg-long, fcd-15} and out of equilibrium \cite{neg-quench-noi, neg-quench}.
Other interesting numerical studies for various one dimensional lattice systems have been performed  in \cite{neg-lattice-1d}.
The same numerical method employed to get the mutual information from the replica limit has been used to get the logarithmic negativity from the replica limit (\ref{replica limit neg}), since similar difficulties occur \cite{rational interpolation-noi}.

In two spatial dimensions, the logarithmic negativity of topological systems has been considered \cite{neg-top} and recently interesting lattice analysis have been performed for both fermionic and bosonic systems \cite{ez-15-2dim, singh-15}.
Some results have been found also in the context of holography \cite{neg-holo}.

In this paper we consider a two dimensional square harmonic lattice with nearest neighbour interactions in its ground state.
We focus on the regime of massless oscillators, whose continuum limit is described by the CFT given by the massless scalar field in $2+1$ dimensions.
In the thermodynamic limit, we study the logarithmic negativity and the quantity (\ref{En def}) for various configurations of adjacent domains
in the regime where they become large. 
At leading order, these quantities follow an area law behaviour in terms of the length of the curve shared by the adjacent regions.
This observation for the logarithmic negativity  has been already done for this model in \cite{ez-15-2dim}, where the configuration given by two halves of a square has been considered.
We notice that the coefficient of the area law term is related to the coefficient of the area law term in the R\'enyi entropies. 
We study also the subleading logarithmic term, which occurs whenever the curve shared by the adjacent regions contains vertices.
Such term is very interesting because it is independent of the regularisation details. 

The layout of this manuscript is as follows. 
In \S\ref{sec:hc} we review the method to compute $S_A$, $S_A^{(n)}$, $\mathcal{E}$ and $\mathcal{E}_n$ for this bosonic lattice.
In \S\ref{sec:area law} we investigate the area law behaviour in the leading term of $\mathcal{E}$ and $\mathcal{E}_n$ for various configurations of large adjacent domains in the infinitely extended lattice. 
In \S\ref{sec:corners} we study the subleading logarithmic term of $\mathcal{E}$ due to the occurrence of vertices in the curve shared by the adjacent domains and in \S\ref{sec:conclusions} we draw some conclusions.


\section{Harmonic lattice}
\label{sec:hc}

In this section we introduce the lattice model considered throughout this manuscript, its correlators in the thermodynamic limit and 
their role in computing the entanglement entropies, the moments of the partial transpose and the logarithmic negativity.

\subsection{Hamiltonian and correlators}

We consider the two dimensional square lattice made by harmonic oscillators coupled through the nearest neighbour spring-like interaction.
Denoting by $L_x$ and $L_y$ the number of sites (oscillators) along the two orthogonal directions, such lattice contains $N=L_x L_y$ oscillators. 
The Hamiltonian of the model  reads
\be
\label{ham}
H= 
\sum_{1\leqslant i \leqslant L_x \atop 1\leqslant j \leqslant L_y }
\bigg\{
\frac{p_{i,j}^2}{2M} + \frac{M \omega^2}{2}\, q^2_{i,j} + \frac{K}{2} 
\Big[ 
\big( q_{i+1, j} - q_{i,j} \big)^2
+
\big( q_{i, j+1} - q_{i,j} \big)^2
\Big]
\bigg\}\,,
\ee
where the pair of integers $(i,j)$ identifies a specific lattice site.
The canonical variables $q_{i,j}$ and $p_{i,j}$ satisfy the canonical commutation relation $[q_{i,j}, q_{r,s}] = [p_{i,j}, p_{r,s}]  = 0$ and $[q_{i,j}, p_{r,s}]  =  \textrm{i}\,\delta_{i,r} \delta_{j,s}$.
We assume periodic boundary conditions along both the spatial directions, namely  $q_{L_x+k, j} = q_{k, j} $, $p_{L_x+k, j} = p_{k, j} $, $q_{j, L_y+k} = q_{j, k} $ and   $p_{j, L_y+k} = p_{j, k} $ for a generic integer $k$.

The model described by (\ref{ham}) contains three parameters $\omega$, $M$ and $K$, but not all of them are independent. 
Indeed, by performing the canonical rescaling $(q_{i,j} , p_{i,j}) \to (\sqrt[4]{MK} q_{i,j} , \, p_{i,j}/\sqrt[4]{MK} )$ and introducing $a=\sqrt{M/K}$, the Hamiltonian (\ref{ham}) becomes
\be
\label{ham2}
H= 
\sum_{1\leqslant i \leqslant L_x \atop 1\leqslant j \leqslant L_y }
\bigg\{
\frac{p_{i,j}^2}{2a} + \frac{a \omega^2}{2}\, q^2_{i,j} + \frac{1}{2a} 
\Big[ 
\big( q_{i+1, j} - q_{i,j} \big)^2
+
\big( q_{i, j+1} - q_{i,j} \big)^2
\Big]
\bigg\}\,.
\ee
From this expression, one can easily observe that \eqref{ham} gives  the Hamiltonian of a free scalar field with mass $\omega$ in two spatial dimensions discretised on a square lattice with lattice spacing $a$.
The continuum limit corresponds to take simultaneously the limits  $L_x \to \infty$, $L_y \to \infty$ and $a\to 0$,  while $L_x a$ and $L_y a$ are kept fixed. 
In our lattice computations, without loss of generality, we set $K=M=1$. 
The Hamiltonian (\ref{ham}) can be diagonalised in a standard way, finding the following dispersion relation
\be
\label{dispersion relation}
\omega_{\boldsymbol{k}}
\equiv
\sqrt{\omega^2 + 4 \Big[ \sin^2(\pi k_x/L_x) + \sin^2(\pi k_y/L_y) \Big]}
\,\geqslant \, \omega \,,
\ee
where $\boldsymbol{k}=(k_x, k_y)$ is a pair of integers such that $0\leqslant k_x < L_x$ and $0\leqslant k_y < L_y$.
Because of the translation invariance of the model, the zero mode with $\boldsymbol{k}=(0, 0)$ occurs, for which the equality holds in (\ref{dispersion relation}).

In our analysis we need the following vacuum correlators 
\bea
\label{qq correlator}
& &\hspace{-2cm}
\langle 0 | q_{i,j} q_{r,s}| 0 \rangle 
\,=\,
\frac{1}{2\, L_x L_y}
\sum_{0\leqslant k_x< L_x \atop 0\leqslant k_y < L_y }
\frac{1}{\omega_{\boldsymbol{k}}}\,
\cos[2\pi k_x (i-r)/L_x]\,
\cos[2\pi k_y (j-s)/L_y]\,,
\\
\label{pp correlator}
\rule{0pt}{.5cm}
& &\hspace{-2cm}
\langle 0|p_{i,j} p_{r,s}|0\rangle 
\,=\,
\frac{1}{2\, L_x L_y}
\sum_{0\leqslant k_x < L_x \atop 0\leqslant k_y < L_y }
\omega_{\boldsymbol{k}}\,
\cos[2\pi k_x (i-r)/L_x]\,
\cos[2\pi k_y (j-s)/L_y]\,,
\eea
which are the matrix elements of the correlation matrices $\mathbb{Q}$ and $\mathbb{P}$ respectively (where $(i,j)$ and $(r,s)$ are the raw and column indices respectively).
These matrices satisfy $\mathbb{Q} \cdot \mathbb{P} = \mathbb{I}/4$, being $\mathbb{I}$ is the identity matrix.
We remark that the term in \eqref{qq correlator} corresponding to the zero mode reads $1/(2 L_x L_y \omega)$, which is divergent for $\omega\to 0$.
This implies that we cannot take $\omega=0$ in a finite lattice.

Since from our computations on the lattice we would like to extract information about the model in the continuum limit, we need to consider the regime where $L_x,L_y \gg \ell \gg 1$, being $\ell$ the linear size of the subsystem. 
Thus, it is convenient to consider the thermodynamic limit, where $L_x\to \infty$ and $L_y \to \infty$, while the lattice spacing $a$ is kept finite. 
In order to perform the thermodynamic limit of the correlators \eqref{qq correlator} and \eqref{pp correlator}, we define $ q_r= 2\pi k_r/L_r$ for $r\in \{x,y\}$. In the thermodynamic limit $q_r$ becomes a continuous variable $q_r \in [0,2\pi)$ and the sum in (\ref{qq correlator}) and (\ref{pp correlator}) is replaced by an integration according to $\frac{1}{L_r}\sum_{k_r}\rightarrow\int_{0}^{2\pi}\, \frac{dq_r}{2 \pi}$.
Thus, the correlators (\ref{qq correlator}) and (\ref{pp correlator}) in the thermodynamic limit become respectively
\bea
\label{qq correlator cont}
& &\hspace{-1cm}
\langle 0 | q_{i,j} q_{r,s}| 0 \rangle 
\,=\,
\frac{1}{8\pi^2}
\int_{0}^{2\pi}
\frac{1}{\omega_{\boldsymbol{q}}}\,
\cos[q_x (i-r)]\,
\cos[q_y (j-s)]
\, dq_x dq_y\,,
\\
\label{pp correlator cont}
\rule{0pt}{.6cm}
& &\hspace{-1cm}
\langle 0|p_{i,j} p_{r,s}|0\rangle 
\,=\,
\frac{1}{8\pi^2}
\int_{0}^{2\pi}
\omega_{\boldsymbol{q}}\,
\cos[q_x (i-r)]\,
\cos[q_y (j-s)]
\, dq_x dq_y\,,
\eea
where $\omega_{\boldsymbol{q}} = \sqrt{\omega^2 + 4 [ \sin^2(q_x/2) + \sin^2(q_y/2)]}$, with $\boldsymbol{q}=(q_x, q_y)$. 
When $\omega =0$ the integral in (\ref{qq correlator cont}) is convergent and therefore, in principle, the massless regime can be considered
without any approximation. Nevertheless, in order to avoid divergent integrands, in our numerical calculations we have set $\omega \leqslant 10^{-6}$, checking in some cases that smaller values of $\omega$ do not lead to significant changes in the final result.

\subsection{Entanglement entropies}

Following \cite{hc-ee-lattice}, we can compute the R\'enyi entropies $S_A^{(n)}$ for this model by considering the matrices  $\mathbb{Q}_A$ and $\mathbb{P}_A$, which are obtained by restricting $\mathbb{Q}$ and $\mathbb{P}$ respectively to the subsystem $A$. 
Their size is $N_A \times N_A$, being $N_A$ the number of lattice points inside the region $A$. 

The matrix product $\mathbb{Q}_A \cdot \mathbb{P}_A$ has positive eigenvalues $\{\mu_1^2,\dots,\mu_{N_A}^2\}$ with $\mu_i^2 \geqslant 1/4$ and the moments of the reduced density matrix are given by 
\be
\label{moments rdm}
\Tr \rho_A^n \,=\,
\prod_{j=1}^{N_A}
\bigg[
\bigg(\mu_j +\frac{1}{2} \bigg)^n - \bigg(\mu_j -\frac{1}{2} \bigg)^n
\,\bigg]^{-1}.
\ee
From this expression it is straightforward to get the R\'enyi entropies
\be
S_A^{(n)} =
 \frac{1}{1-n}\, \log \Tr \rho_A^n 
\,=\,
\frac{1}{n-1}\,
\sum_{j=1}^{N_A} \, \log \bigg[
\bigg(\mu_j +\frac{1}{2} \bigg)^n - \bigg(\mu_j -\frac{1}{2} \bigg)^n
\,\bigg]\,,
\ee
while the entanglement entropy is given by
\be
S_A \,=\,
\sum_{j=1}^{N_A} 
\bigg[
\bigg(\mu_j +\frac{1}{2} \bigg) \,\log \bigg(\mu_j +\frac{1}{2} \bigg)
-
\bigg(\mu_j -\frac{1}{2} \bigg) \, \log \bigg(\mu_j -\frac{1}{2} \bigg)
\bigg]\,.
\ee

By employing these formulas for disjoint domains on the lattice, one gets $I_{A_1,A_2}^{(n)}$ and the mutual information $I_{A_1,A_2}$.

\subsection{Moments of the partial transpose and logarithmic negativity}

In \cite{neg-hc-momentum} it was shown that the partial transposition with respect to $A_2$ for a bosonic state corresponds to the time reversal applied only to the degrees of freedom in $A_2$, while the remaining ones are untouched. 
In particular, in $A_2$ the positions are left invariant $q_{i,j} \to q_{i,j} $ everywhere, while the momenta are reversed $p_{i,j}\to -p_{i,j}$ if ${(i,j) \in A_2}$.
Given a bosonic Gaussian state, like the ground state of the harmonic chain we are considering, the resulting operator after such transformation will be Gaussian as well. 
It is worth remarking that this is not true for fermionic systems. 
For instance, for free fermions the partial transpose of the ground state density matrix can be written as a sum of two Gaussian operators \cite{ez-15}.

The above observations are implemented on our lattice model by introducing the following matrix
\be
\mathbb{P}_{A}^{T_2} = \mathbb{R}_{A_2} \cdot \mathbb{P}_A \cdot \mathbb{R}_{A_2}\,,
\ee
where $\mathbb{R}_{A_2}$ is the $N_A\times N_A$ diagonal matrix having $-1$ in correspondence of the sites belonging to $A_2$ and $+1$ otherwise. 
Since $\mathbb{R}_{A_1}=-\mathbb{R}_{A_2}$, it is easy to observe that $\mathbb{P}_{A}^{T_1}=\mathbb{P}_{A}^{T_2} $, as expected.
The matrix  $\mathbb{Q}_A \cdot \mathbb{P}_A^{T_2}$ has also positive eigenvalues $\{\nu_1^2,\dots, \nu_{N_A}^2\}$, but if the state is entangled some of them can be smaller than $1/4$.
From the eigenvalues $\nu_j$ one gets the moments of the partial transpose of the reduced density matrix as in (\ref{moments rdm}) for the moments of the reduced density matrix, namely
\be
\Tr \big(\rho_A^{T_2}\big)^n \,=\,
\prod_{j=1}^{N_A}
\bigg[
\bigg(\nu_j +\frac{1}{2} \bigg)^n - \bigg(\nu_j -\frac{1}{2} \bigg)^n
\,\bigg]^{-1}.
\ee
The trace norm of $\rho_A^{T_2}$ reads
\be
\big\| \rho_A^{T_2} \big\|
\,=\,
\prod_{j=1}^{N_A}
\bigg[\,
\bigg|\nu_j +\frac{1}{2} \bigg| - \bigg|\nu_j -\frac{1}{2} \bigg|
\,\bigg]^{-1}
\,=\,
\prod_{j=1}^{N_A}
\textrm{max}\bigg(1, \frac{1}{2\nu_j}\bigg) \,,
\ee
which leads straightforwardly to the logarithmic negativity 
\be
\label{eq:neg lattice}
\mathcal{E} = \sum_{j=1}^{N_A} \, \log \big[ \textrm{max}\big(1, (2\nu_j)^{-1}\big)\big] \,.
\ee

In the remaining part of the manuscript we discuss the numerical results obtained by employing the above lattice formulas.


\section{Area law}
\label{sec:area law}

In this section we consider various configurations of adjacent domains for the harmonic lattice in the thermodynamic limit described in \S\ref{sec:hc}.
For large domains, we show that at leading order the logarithmic negativity and $\mathcal{E}_n$ satisfy an area law in terms of the length of the curve shared 
by the adjacent domains. 
We observe that the coefficient of such term in these quantities is related to the coefficient of the area law term in the R\'enyi entropies.

\subsection{Logarithmic negativity}
\label{sec: area log neg}

\begin{figure}
\vspace{.2cm}
\hspace{-1.4cm}
\includegraphics[width=1.15\textwidth]{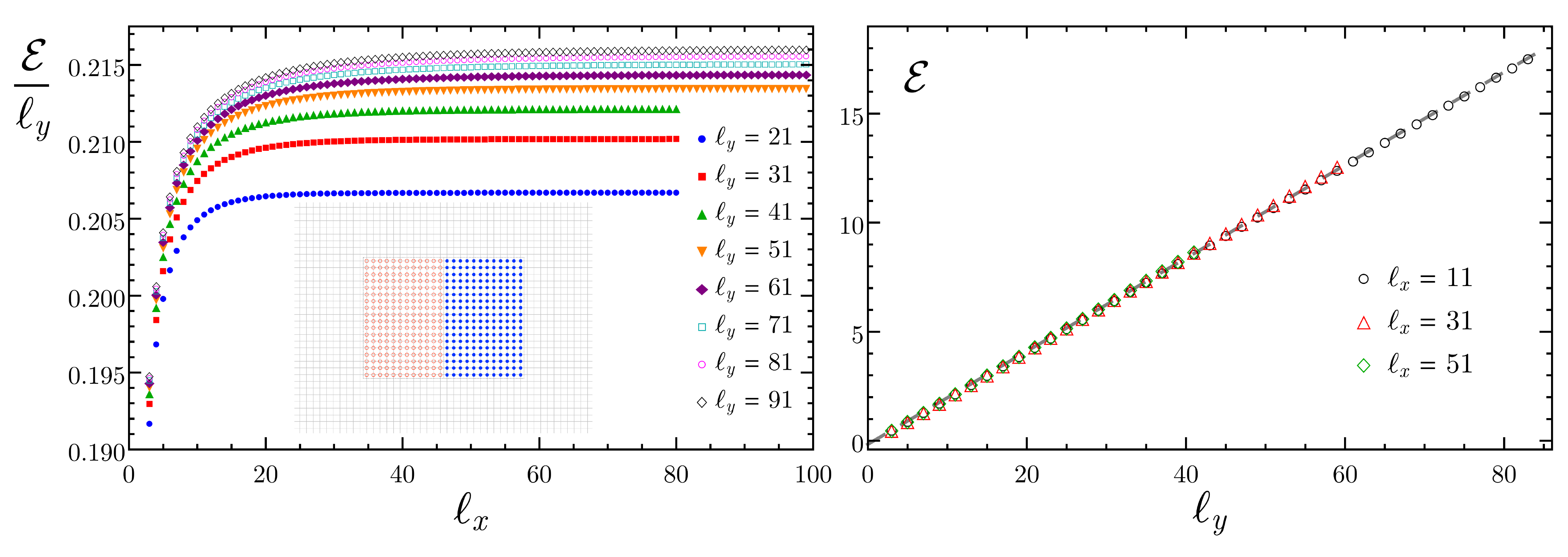}
\vspace{-.7cm}
\caption{Area law behaviour for  the logarithmic negativity $\mathcal{E}$ between two equal rectangles, whose edges have lengths $\ell_x$ and $\ell_y$, which are adjacent along the vertical edge (inset of the left panel).
Left: For fixed values of $\ell_y$, the ratio $\mathcal{E}/\ell_y$ reaches a constant value as  $\ell_x$ increases.
Right: For fixed and large enough values of $\ell_x$, the logarithmic negativity grows linearly as $\ell_y$ increases
(the dashed line is obtained by fitting all the data corresponding to $\ell_x=11$).
}
\label{fig:AdjRect}
\end{figure}

We begin our analysis by considering the logarithmic negativity of two equal adjacent rectangles $A_1$ and $A_2$ which share an edge along the vertical $y$ axis,  
as shown in the inset of the left panel in Fig.\,\ref{fig:AdjRect}, where the adjacent domains are highlighted by blue dots and red circles. 
These rectangles have the natural orientation induced by the underlying lattice, namely their edges are parallel to the vectors generating the square lattice.
Denoting by $\ell_x$ and $\ell_y$ the lengths of the edges along the $x$ and $y$ directions respectively, the numerical data for $\mathcal{E}$ of this configuration 
of adjacent domains are plotted in Fig.\,\ref{fig:AdjRect}.

In the left panel we show the ratio $\mathcal{E}/\ell_y$ as function of $\ell_x$  when $\ell_y$ is kept fixed.
For any given $\ell_y$, such ratio reaches a constant value when $\ell_x$ is sufficiently large.
This confirms the intuition that the main contribution to a quantity characterising the entanglement between two adjacent regions $A_1$ and $A_2$ should come from the degrees of freedom localized along their shared boundary, namely the curve $\partial A_1 \cap \partial A_2$.
In the right panel of Fig.\,\ref{fig:AdjRect}, the logarithmic negativity of the same configuration is plotted as function of $\ell_y$ for fixed values of $\ell_x$.
If $\ell_x$ is sufficiently large, a neat linear growth can be observed. 
The fact that the asymptotic value of  $\mathcal{E}/\ell_y$ depends on  $\ell_y$ in the left panel of Fig.\,\ref{fig:AdjRect} is mainly due to the subleading corner contributions, which will be largely discussed in \S\ref{sec:corners}.

These results tell us that, given two equal and large enough adjacent regions $A_1$ and $A_2$, at leading order the logarithmic negativity $\mathcal{E}$ between them increases like the length of the curve $\partial A_1 \cap \partial A_2$ shared by their boundaries as their size increases.
Such length will be denoted by $P_{\textrm{\tiny shared}} \equiv \textrm{length}(\partial A_1 \cap \partial A_2)$ throughout this manuscript.
Thus, the logarithmic negativity between large adjacent domains satisfies an area law in terms of the region shared by their boundaries. 
This observation has been recently done for this model also by Eisler and Zimbor\'as \cite{ez-15-2dim}, who have considered the logarithmic negativity between the two halves of a square
as the length of its edge increases.

\begin{figure}[t!]
\vspace{.1cm}
\hspace{-.7cm}
\includegraphics[width=1.08\textwidth]{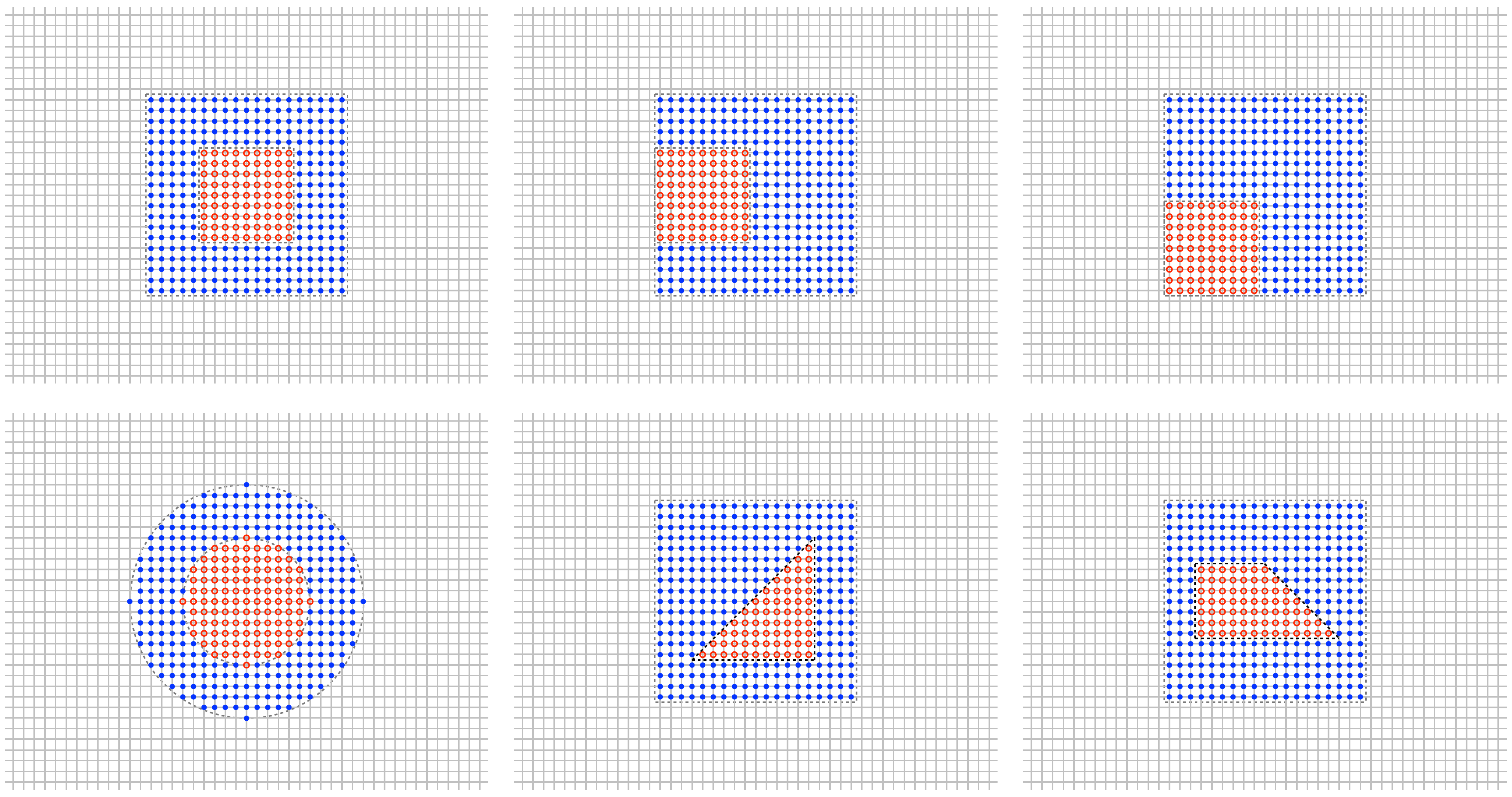}
\vspace{-.4cm}
\caption{Configurations of adjacent domains on the lattice, identified by red circles and blue dots, which have been employed to study the area law behaviour (see \S\ref{sec:area law} and Figs.\,\ref{fig:NegArea} and \ref{fig:renyi}) and the corner contributions for explementary angles (see \S\ref{sec:corners} and Fig.\,\ref{fig:cornerNegOld}).
}
\label{fig:configs1}
\end{figure}

In order to improve our analysis of the area law behaviour for the logarithmic negativity between adjacent regions $A_1$ and $A_2$, let us consider the six configurations of adjacent domains on the lattice shown in Fig.\,\ref{fig:configs1}, where the sites belonging to $A_1$ and $A_2$ are highlighted by blue dots and red circles. 
In these examples the curve $\partial A_1 \cap \partial A_2$ is not given by a simple line segment.
The domains identified by the red circles in Fig.\,\ref{fig:configs1} are convex, while the ones corresponding to the blue dots are not.

It is well known that the curve separating adjacent domains on the lattice is not unique.
For these configurations we have chosen the dashed lines, which are the lines whose length has been used to get the perimeter. 
The three configurations in the top panels of Fig.\,\ref{fig:configs1} are natural to define on the square lattice because their edges are parallel to the orthogonal vectors generating the lattice.
Instead, the three configurations in the bottom panels of Fig.\,\ref{fig:configs1} are made by adjacent domains  where the line $\partial A_1 \cap \partial A_2$ either is curved or it contains a line segment which is oblique with respect to the vectors generating the lattice.
Notice that a disk of given radius on the lattice could include a different number of sites depending on whether the centre of the disk is located on a lattice site or within a plaquette.
Such ambiguity does not affect the leading order behaviour of the quantities that we are considering, but it could be relevant for subleading terms \cite{ch-review, CasiniMyers-15 FthMI}.

Also for the logarithmic negativity of the adjacent domains shown in Fig.\,\ref{fig:configs1} we have observed the same qualitative behaviour described in the left panel of Fig.\,\ref{fig:AdjRect} for the equal adjacent rectangles: by keeping fixed the region corresponding to the red circles while the sizes of the region characterised by the blue dots increase, $\mathcal{E}$ saturates to a constant value.

\begin{figure}[t!]
\vspace{.1cm}
\hspace{-1.2cm}
\includegraphics[width=1.15\textwidth]{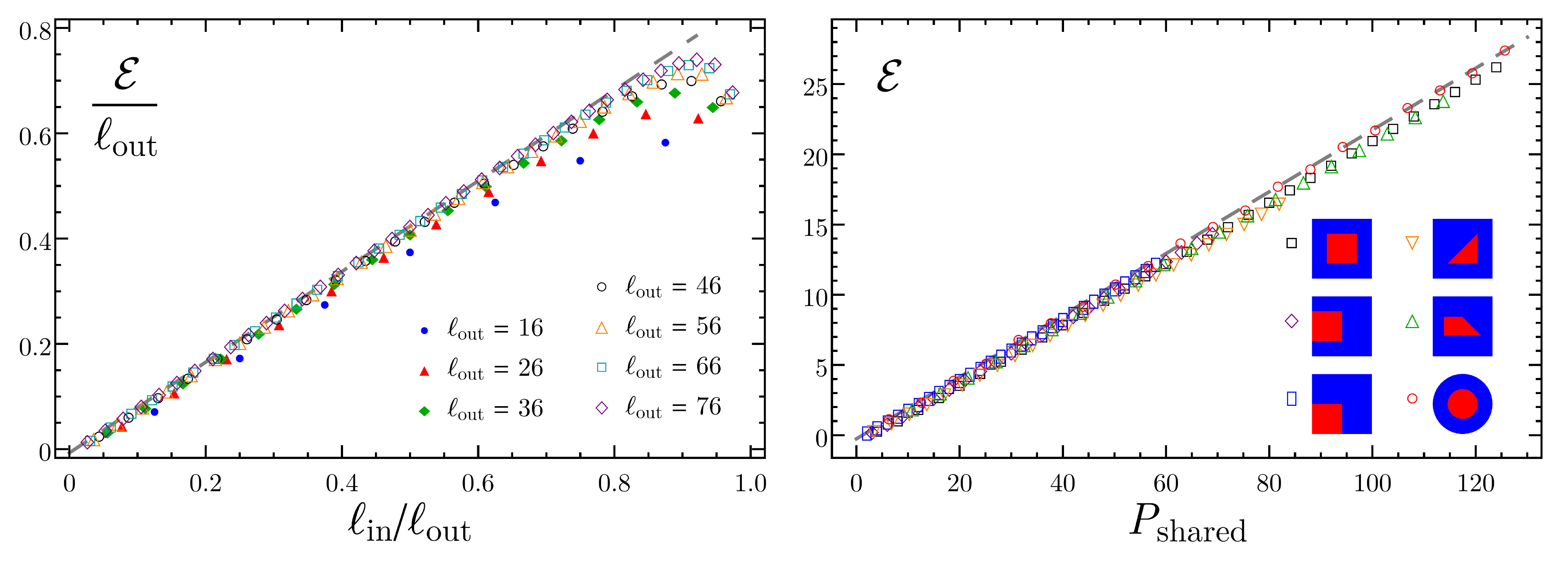}
\vspace{-.6cm}
\caption{Left: Logarithmic negativity for the configuration in the top left panel of Fig.\,\ref{fig:configs1}, where $\ell_{\textrm{\tiny in }}$ and $\ell_{\textrm{\tiny out}}$ are the sizes of the edges of the red square and of the square $A_1 \cup A_2$ respectively. 
The dashed line is obtained by fitting the data corresponding to $\ell_{\textrm{\tiny out}} =76$ up to $\ell_{\textrm{\tiny in}}  / \ell_{\textrm{\tiny out}} \simeq 0.7$.
Right: Logarithmic negativity of adjacent domains for the configurations shown in Fig.\,\ref{fig:configs1}, which involve different shapes for the curve $\partial A_1 \cap \partial A_2$.
The sizes of the domains increase while their ratios are kept fixed.
The data correspond to configurations where the linear size of the convex domains (highlighted by red circles) is $1/3$ of the size of the corresponding $A_1 \cup A_2$.
The dashed line has been found by fitting the data obtained for the configuration in the bottom left panel of Fig.\,\ref{fig:configs1}.
}
\label{fig:NegArea}
\end{figure}

In Fig.\,\ref{fig:NegArea} we show some quantitative results for the logarithmic negativity of the configurations in Fig.\,\ref{fig:configs1}.
In particular, let us consider the configuration in the top left panel, which is characterised by the lengths $\ell_{\textrm{\tiny in}}<\ell_{\textrm{\tiny out}}$ of the edges of the internal square and of the whole subregion $A_1 \cup A_2$ respectively.
In the left panel of Fig.\,\ref{fig:NegArea} we show $\mathcal{E}/\ell_{\textrm{\tiny out}}$ as function of the ratio $\ell_{\textrm{\tiny in}} / \ell_{\textrm{\tiny out}} < 1$ when $\ell_{\textrm{\tiny out}}$ is kept fixed and the internal square increases.
For large enough $\ell_{\textrm{\tiny out}}$, the area law behaviour in terms of $\ell_{\textrm{\tiny in}}$ is observed.
It is worth remarking that  $\mathcal{E} \to 0$ when $\ell_{\textrm{\tiny in}}/ \ell_{\textrm{\tiny out}} \to 1$.
This is expected because in this limit the internal convex domain becomes the whole $A$.
For any fixed and large enough value of $\ell_\text{\tiny out}$ there is a critical size of the internal square after which the logarithmic negativity deviates from the linear growth predicted by the area law. 
The numerical data tell us that such critical value of $\ell_\text{\tiny in}$ increases by increasing $\ell_\text{\tiny out}$.
This suggests that in the continuum limit, where both $\ell_\text{\tiny in}$ and $\ell_\text{\tiny out}$ diverge but their ratio is finite, the linear behaviour occurs for any ratio $\ell_\text{\tiny in}/\ell_\text{\tiny out}<1$.
From the plot in the left panel of Fig.\,\ref{fig:NegArea} one can also notice that the ratio $ \ell_{\textrm{\tiny in}} / \ell_{\textrm{\tiny out}} \simeq 1/3$ is a good regime to explore the area law behaviour even for relatively small domains. 
Given the latter observation, we have considered the logarithmic negativity of all the configurations in Fig.\,\ref{fig:configs1} with a ratio of $1/3$ between the size of the internal convex domain and the one of the whole subsystem $A$ (for the triangle we refer to its shortest edge and for the trapezoid to its height).
By increasing the sizes of both the domains while their ratio is kept fixed to $1/3$, we find the results collected in the right panel of Fig.\,\ref{fig:NegArea}, which nicely confirm the area law behaviour in terms of $P_{\textrm{\tiny shared}} $ observed above. 
Notice that the different configurations in the right panel of Fig.\,\ref{fig:NegArea} provide linear growths with almost the same slope.
Moreover, the data corresponding to the configuration in the bottom left panel of Fig.\,\ref{fig:configs1} do not provide a neat straight line, as expected whenever domains with a curved boundary on a square lattice are involved.

Summarising the numerical results presented above, we can conclude that at the leading order the logarithmic negativity of two large adjacent domains  $A_1$ and $A_2$ on a lattice of massless harmonic oscillators with nearest neighbour spring-like interactions  in the ground state satisfies an area law in terms of the length $P_{\textrm{\tiny shared}}$ of the curve shared by the adjacent regions, namely
\be
\label{area law neg}
\mathcal{E} \,=\, a\, P_{\textrm{\tiny shared}} + \dots\,,
\ee
where the dots indicate subleading terms for large values of $P_{\textrm{\tiny shared}}$.
The area law (\ref{area law neg}) is consistent with the fact that $\mathcal{E}$ measures the bipartite entanglement between $A_1$ and $A_2$ for the mixed state characterised by the reduced density matrix $\rho_{A_1 \cup A_2}$.
The coefficient $a$ in (\ref{area law neg}) is non universal, i.e. it depends on the ultraviolet details.

Given the two adjacent regions $A_1$ and $A_2$ considered above, another very interesting quantity to study is their mutual information $I_{A_1, A_2} $, which has been defined in (\ref{MI def}).
Since the area law of the entanglement entropy for large domains tells us that $S_A = \tilde{a} \,P_A + \dots$, it is straightforward to find that $I_{A_1, A_2} $ of adjacent domains satisfies an area law in terms of $P_{\textrm{\tiny shared}}$. In particular, we have that
\be
\label{area law MI}
I_{A_1, A_2} \,=\, 2\tilde{a}\, P_{\textrm{\tiny shared}} + \dots\,,
\ee
where, as above, the dots stand for subleading terms.

\subsection{Moments of the partial transpose}
\label{sec: area neg renyi}

The moments $\Tr (\rho_A^{T_2})^n$ of the partial transpose for integer values of $n$ are interesting quantities to study because they provide the logarithmic negativity through the replica limit (\ref{replica limit neg}) \cite{cct-neg-letter,cct-neg-long}.

Given the configurations of adjacent domains described in \S\ref{sec: area log neg}, instead of considering the $n$-th moment of the partial transpose, we find it more interesting the ratio $\mathcal{E}_n$ defined in (\ref{En def}), 
which also provides the logarithmic negativity through the replica limit (\ref{replica limit E_n}) because of the normalisation condition $\Tr \rho_A = 1$.
In our model, the main reason to consider $\mathcal{E}_n$ instead of $\log \Tr (\rho^{T_2} )^{n}$ occurring in (\ref{replica limit neg}) is that, by repeating the analysis described in \S\ref{sec: area log neg}, we find that, at leading order for large adjacent domains,  $\mathcal{E}_n$ follows an area law in terms of the length of the curve shared by the adjacent domains, i.e.
\be
\label{area law renyi neg}
\mathcal{E}_n \,=\, a_n \, P_{\textrm{\tiny shared}} + \dots\,,
\ee
where the non universal coefficient $a_n$ depends on the integer $n$ and the dots denote subleading terms. 
We recall that the R\'enyi entropies of our model satisfy the area law $S_A^{(n)} = \tilde{a}_n \,P_A + \dots $, where the coefficient $\tilde{a}_n$ is non universal as well and $\lim_{n\to 1}  \tilde{a}_n=  \tilde{a}$.

From (\ref{area law renyi neg}) and the area law of the R\'enyi entropies it is straightforward to find the leading term of the logarithm of the moments of the partial transpose, which is given by 
\begin{equation}
\log \Tr \big( \rho_A^{T_2} \big)^n =\, 
 a_n \, P_\textrm{\tiny shared} + (1-n)\tilde a_n \, P_A+\dots\,.
\end{equation}
Thus, the quantities $\log \Tr ( \rho_A^{T_2} )^n $ contain an area law contribution also from the boundary of $A=A_1 \cup A_2$.
Since  $\lim_{n_e \rightarrow 1} (n_e-1)\tilde a_{n_e} = 0$, such term cancels in the replica limit (\ref{replica limit neg}).
A similar cancellation occurs also for adjacent intervals in  $1+1$ dimensional CFTs.
Indeed, considering the divergent terms for $\varepsilon \to 0$, in $\mathcal{E}_n$ only the ones giving a non trivial contribution after the replica limit survive, while $\log \Tr (\rho^{T_2} )^{n}$ contains also other terms \cite{cct-neg-long}, which vanish in the replica limit (\ref{replica limit neg}).
 The quantity $\mathcal{E}_n$ for adjacent intervals in $1+1$ dimensional CFTs  has been studied in \cite{cct-neg-T, neg-quench-noi} and for free fermions on a two dimensional lattice in \cite{ez-15-2dim}.

\begin{figure}[t!]
\vspace{.1cm}
\hspace{-1.4cm}
\includegraphics[width=1.15\textwidth]{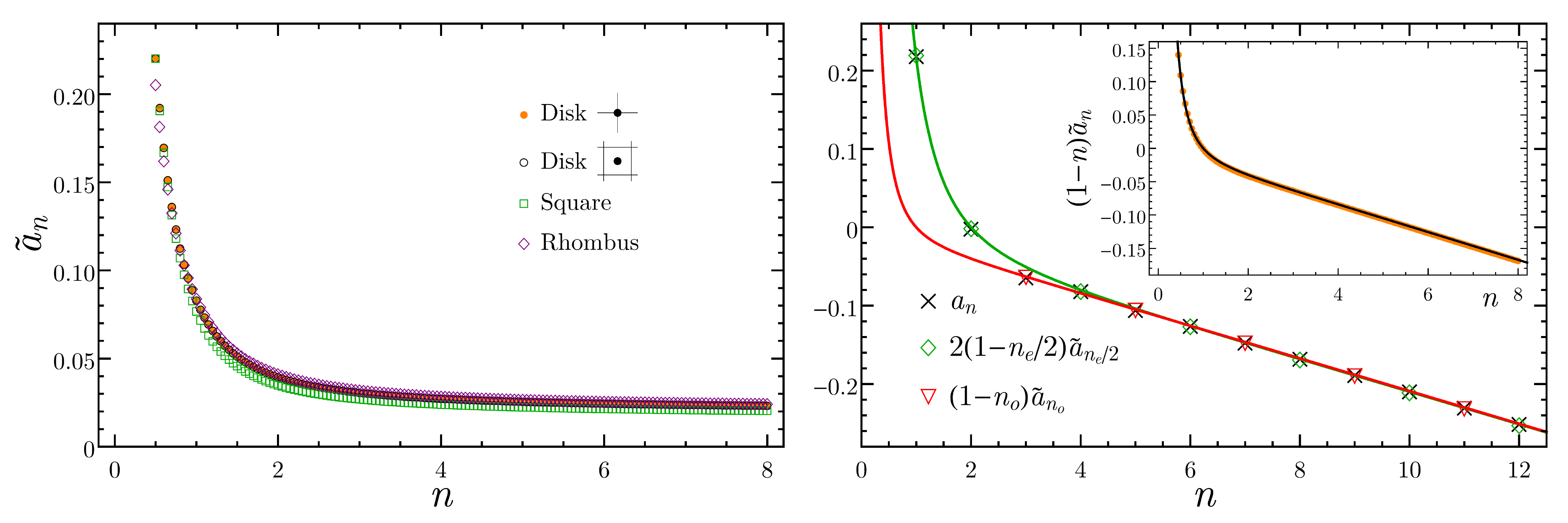}
\vspace{-.6cm}
\caption{Left: The coefficient of the area law term in the R\'enyi entropies $S_A^{(n)}$ as function of $n$ for domains with different shapes. centre
The centres of the disks have been chosen both on the lattice site and in the centre of a plaquette.
Here the edges of the square configuration are parallel to the vectors generating the lattice, while the rhombus configuration corresponds
to the previous square configuration rotated by $\pi/4$ with respect to its centre. 
Right: Numerical check of (\ref{area law coeff a_n}) and (\ref{area coeff neg renyi 1/2}) for the configuration in the bottom left panel of Fig.\,\ref{fig:configs1} with $R_{\textrm{\tiny in}} / R_{\textrm{\tiny out}} =1/3$. 
Inset: The coefficient $(1-n)\tilde{a}_n$ as function of $n$ for disks (the data are taken from the left panel with the same colour code). 
The black curve corresponds to the best fit of the numerical data through the function $f(n)=c_{-2}/n^2 + c_{-1}/n + c_0 + c_1 n $, 
 constrained by the condition $f(1)=0$.
In the main plot the red curve is $f(n)$ obtained in the inset and the green one is $2f(n/2)$.
}
\label{fig:renyi}
\end{figure}

In the left panel of Fig.\,\ref{fig:renyi} we show $\tilde{a}_n$ as function of $n$ for disks and squares on the infinite lattice.
The centres of the disks have been chosen either on a lattice site (like in the bottom left panel of Fig.\,\ref{fig:configs1}) or in the central point of a plaquette.
As for the squares, we have considered both the ones whose edges are parallel to the vectors generating the lattice and the ones obtained by rotating of $\pi/4$ the previous ones (denoted as rhombi in the plot).
In this plot we have $150 \leqslant P_{\textrm{\tiny shared}}  \leqslant  200$, depending on the configuration. 
A slight dependence of $\tilde{a}_n$ on the shape can be observed from our data points. 
The asymptotic $\tilde{a}_n \sim 1/n^2$ as $n\to 0$ \cite{cft-n=0} is consistent with our numerical results. 
For the Ising model a numerical analysis for $\tilde{a}_n$ as  $n\to 0$  has been done in \cite{gt-09}.
The numerical data in the left panel of Fig.\,\ref{fig:renyi} have been found by employing a fitting function which includes also a logarithmic term, as it will be discussed in detail in \S\ref{sec:corners},
but such term does not change the coefficient of the leading area law term in a significant way.

It is worth considering the quantity $I_{A_1,A_2}^{(n)}=S_{A_1}^{(n)} + S_{A_2}^{(n)} - S_{A_1 \cup A_2}^{(n)} $ for the configurations of adjacent domains described in \S\ref{sec: area log neg}.
Given the area law behaviour of $S_{A}^{(n)}$, it is easy to observe that $I_{A_1,A_2}^{(n)}$ displays an area law behaviour in terms of $P_{\textrm{\tiny shared}}$, namely
\be
\label{area law renyi MI}
I^{(n)}_{A_1, A_2} \,=\, 2\tilde{a}_n\, P_{\textrm{\tiny shared}} + \dots\,,
\ee

Once the configuration of adjacent domains $A_1$ and $A_2$ has been chosen, we find it interesting to compare the non universal coefficients $a_n $ and $\tilde{a}_n $ occurring in the area law terms of (\ref{area law renyi neg}) and  (\ref{area law renyi MI}) respectively.

It is reasonable to expect that the area law term for $\mathcal{E}_n$ comes from effects localised in the neighbourhood of the curve $\partial A_1 \cap \partial A_2$.
Thus, considering e.g. the configurations in Fig.\,\ref{fig:configs1} where the domain identified by the red circles is entirely surrounded by the one characterised by the blue dots (i.e. the configurations in the top left, bottom middle and bottom right panels), such term should be independent of the size of the domain identified by the blue dots. 
In the limit where this domain becomes  the whole region complementary to the one identified by the red circles, one gets the bipartition of the ground state. 
These considerations suggest us that $a_n$ in (\ref{area law renyi neg}) is the same one occurring for a bipartition of the ground state, when the identity (\ref{identity pure states}) can be applied. 
This implies that the following relation should hold
\be
\label{area law coeff a_n}
a_n =  \Bigg\{\begin{array}{ll}
\big(1-n_o\big)\, \tilde{a}_{n_o}   & \hspace{.5cm} \textrm{odd $n=n_o$}\,,
\\
\rule{0pt}{.6cm}
2 \big(1-n_e/2\big)\,\tilde{a}_{n_e/2}  &  \hspace{.5cm} \textrm{even $n=n_e$}\,.
\end{array}
\ee

Notice that $a_2=0$, as expected.
By employing the relation (\ref{area law coeff a_n}) and the replica limit (\ref{replica limit E_n}), 
it is straightforward to find that the coefficient of the area law term in the logarithmic negativity in (\ref{area law neg}) is equal to the 
coefficient of the area law term in the R\'enyi entropy of order $1/2$, namely
\be
\label{area coeff neg renyi 1/2}
a \,=\,\tilde{a}_{1/2}\,.
\ee

In the right panel of Fig.\,\ref{fig:renyi} we show a numerical check of the relations (\ref{area law coeff a_n}) and (\ref{area coeff neg renyi 1/2}) for the configuration in the bottom left panel of Fig.\,\ref{fig:configs1}.
In particular, the coincidence of the data points corresponding to $n=1/2$ provides a check of (\ref{area coeff neg renyi 1/2}).
The solid curve in the inset is obtained by fitting the data with the function $f(n) = c_{-2}/n^2 + c_{-1}/n + c_0 + c_1 n $, 
where the parameters are constrained by the requirement that $f(1)=0$.
Thus, such fit has three independent parameters.
As for the solid curves in the main plot, the red one is $f(n)$, namely the black curve found in the inset, while the green one is $2f(n/2)$.

This analysis has been performed also for other configurations as further checks of (\ref{area law coeff a_n}) and (\ref{area coeff neg renyi 1/2}), finding the same qualitative behaviours.

\section{Logarithmic term from the corner contributions}
\label{sec:corners}

In this section we consider the subleading logarithmic term in $\mathcal{E}$ and $\mathcal{E}_n$ for adjacent domains, which occurs whenever 
the shared curve $\partial A_1 \cap \partial A_2$ contains some vertices, where its endpoints are included among them.
For vertices corresponding to explementary angles, we provide some numerical evidence that the corner function of $\mathcal{E}$ is given by the corner
function of the R\'enyi entropy of order $1/2$.

\subsection{Entanglement entropies}
\label{sec:corners ent}

Let us consider the entanglement entropy $S_A$ of a connected domain $A$ whose boundary contains some vertices (see Figs.\,\ref{fig:configs1}, \ref{fig:configs2} and \ref{fig:configs3} for examples). 
For large size of $A$, the leading term gives the area law behaviour. 
The occurrence of vertices in $\partial A$ provides a subleading logarithmic term which is characterised by a corner function $\tilde{b}(\theta) $ as follows  \cite{ch-06-corners, chl-08-corners, ch-review}
\be
\label{S_A corner}
S_A = \tilde{a}\, P_A  - \bigg(
\sum_{{\textrm{vertices}} \atop {\textrm{of $\partial A$}}}
\tilde{b}(\theta_i)  
\bigg) \log P_A + \dots
\qquad
0< \,\theta_i \leqslant \pi\,,
\ee
where $\theta_i$ is the opening angle in $A$ corresponding to the $i$-th vertex of $\partial A$ and the dots denote subleading terms.

Since the logarithmic term is due to the corners, we have that $\tilde{b}(\pi) = 0 $.
From the general property that $S_A= S_B$ for pure states and a bipartite Hilbert space $\mathcal{H} = \mathcal{H}_A \otimes \mathcal{H}_B$, we have that $\tilde{b}(\theta) = \tilde{b}(2\pi-\theta)$, which tells us that $\tilde{b}(\theta)$ is defined for $0< \theta \leqslant \pi$.
The model dependent corner function $\tilde{b}(\theta)$ is universal, i.e. independent of the ultraviolet details of the regularisation.
In the continuum limit, which is described by a $2+1$ dimensional CFT, the corner function $\tilde{b}(\theta)$ contains important information about the model.
For instance, recently it has been found that the constant $\sigma$ entering in the asymptotic behaviour $\tilde{b}(\theta) = \sigma (\pi-\theta)^2 + \dots $ as $\theta \to \pi$ at the leading order is related to the constant characterising the correlator $\langle T_{\mu\nu}(x) T_{\alpha\beta}(y)  \rangle $ of the underlying CFT \cite{corner-recent qft}.
The corner function $\tilde{b}(\theta)$ for the massless scalar has been studied by Casini and Huerta \cite{ch-06-corners}.
In the context of holography, by employing the prescription of \cite{RT} for $S_A$, the corner function $\tilde{b}(\theta)$ has been studied e.g. in \cite{holography-corner}.

We find it worth considering the mutual information (\ref{MI def}) of two adjacent domains $A_1$ and $A_2$ when their boundaries contain some vertices.
For the sake of simplicity, we focus on configurations such that either two or three curves meet at every vertex. 
Explicit examples are shown in Figs.\,\ref{fig:configs1}, \ref{fig:configs2} and \ref{fig:configs3}, where the adjacent domains are identified by blue dots and red circles. 
In the scaling limit, when their sizes increase while the ratios among them are kept fixed, from (\ref{S_A corner}) one finds that
\be
\label{area law MI corners}
I_{A_1, A_2} \,=\, 
2\tilde{a}\, P_{\textrm{\tiny shared}} 
- \bigg(
\sum_{{\textrm{vertices of}} \atop {\textrm{$\partial A_1 \cap \partial A_2$}}} 
\hspace{-.15cm}
\Big[ \,
\tilde{b}(\theta^{(1)}_i) + \tilde{b}(\theta^{(2)}_i)  - \tilde{b}(\theta^{(1 \cup 2)}_i) 
\,\Big]
\bigg)
\log \ell + \dots \,,
\ee
where $\ell$ is a parameter characterising the common size of the adjacent regions, $\theta^{(k)}_i$ is the angle in $A_k$  
and $\theta_i^{(1 \cup 2)}$ the angle in $A_1 \cup A_2$ corresponding to the $i$-th vertex belonging to $\partial A_1 \cap \partial A_2$.
In the simplest case where such vertex is not an endpoint of $\partial A_1 \cap \partial A_2$, it provides a bipartition of the angle of $2\pi$ and the corresponding contribution to the sum in (\ref{area law MI corners}) is $2 \tilde{b}(\theta^{(1)}_i)  = 2 \tilde{b}(\theta^{(2)}_i) $.

We find it instructive to consider explicitly the examples in Figs.\,\ref{fig:configs1} and \ref{fig:configs2}.
For  the configurations shown in the top left, middle and right panel of Fig.\,\ref{fig:configs1}, the sum within the parenthesis multiplying  the logarithmic term in (\ref{area law MI corners}) is given by  $8\tilde{b}(\pi/2)$, $8\tilde{b}(\pi/2)$  and $6\tilde{b}(\pi/2)$ respectively, while for the bottom middle and right panels of the same figure it reads $4\tilde{b}(\pi/4) + 2\tilde{b}(\pi/2)$ and $2\tilde{b}(\pi/4) + 4\tilde{b}(\pi/2) + 2\tilde{b}(3\pi/4)$ respectively.
As for the configurations of Fig.\,\ref{fig:configs2}, such coefficient is  $4\tilde{b}(\pi/4)-2\tilde{b}(\pi/2)$, $3\tilde{b}(\pi/4)+ \tilde{b}(3\pi/4)- \tilde{b}(\pi/2)$ and $4\tilde{b}(\pi/2)$ for the top left, middle and right panels respectively,
while it is given by  $3\tilde{b}(\pi/4)+ \tilde{b}(3\pi/4)-2\tilde{b}(\pi/2)$, $2\tilde{b}(\pi/2)$ and $2\tilde{b}(\pi/4)+ 2\tilde{b}(3\pi/4)-2\tilde{b}(\pi/2)$ for the bottom left, middle and right panels respectively.
It is not difficult to get the coefficient of the logarithmic term also for the configurations in Fig.\,\ref{fig:configs3}. 
Let us point out that for the one in the top left panel such coefficient is vanishing.

As for the R\'enyi entropies of domains with non smooth boundary, we have that
\be
\label{renyi corner}
S^{(n)}_A = \tilde{a}_n \, P_A  - 
\bigg(
\sum_{{\textrm{vertices}} \atop {\textrm{of $\partial A$}}}
\tilde{b}_n(\theta_i)  
\bigg) \log P_A + \dots \,,
\ee
where the corner function $\tilde{b}_n(\theta) $ depends on the order $n$ and it provides $\tilde{b}(\theta) $ through the replica limit, i.e. $\lim_{n \to 1} \tilde{b}_n(\theta_i)  = \tilde{b}(\theta_i)  $.
For the model we are dealing with, the corner function $\tilde{b}_n(\theta) $ has been found in \cite{ch-06-corners}.
A formula similar to (\ref{area law MI corners}) can be written for $I_{A_1, A_2}^{(n)} $, by just replacing $\tilde{a}$ with  $\tilde{a}_n$ and $\tilde{b}(\theta)$ with  $\tilde{b}_n(\theta)$.

\subsection{Logarithmic negativity}
\label{sec:corner logneg}

\begin{figure}[t!]
\vspace{.1cm}
\hspace{-.7cm}
\includegraphics[width=1.08\textwidth]{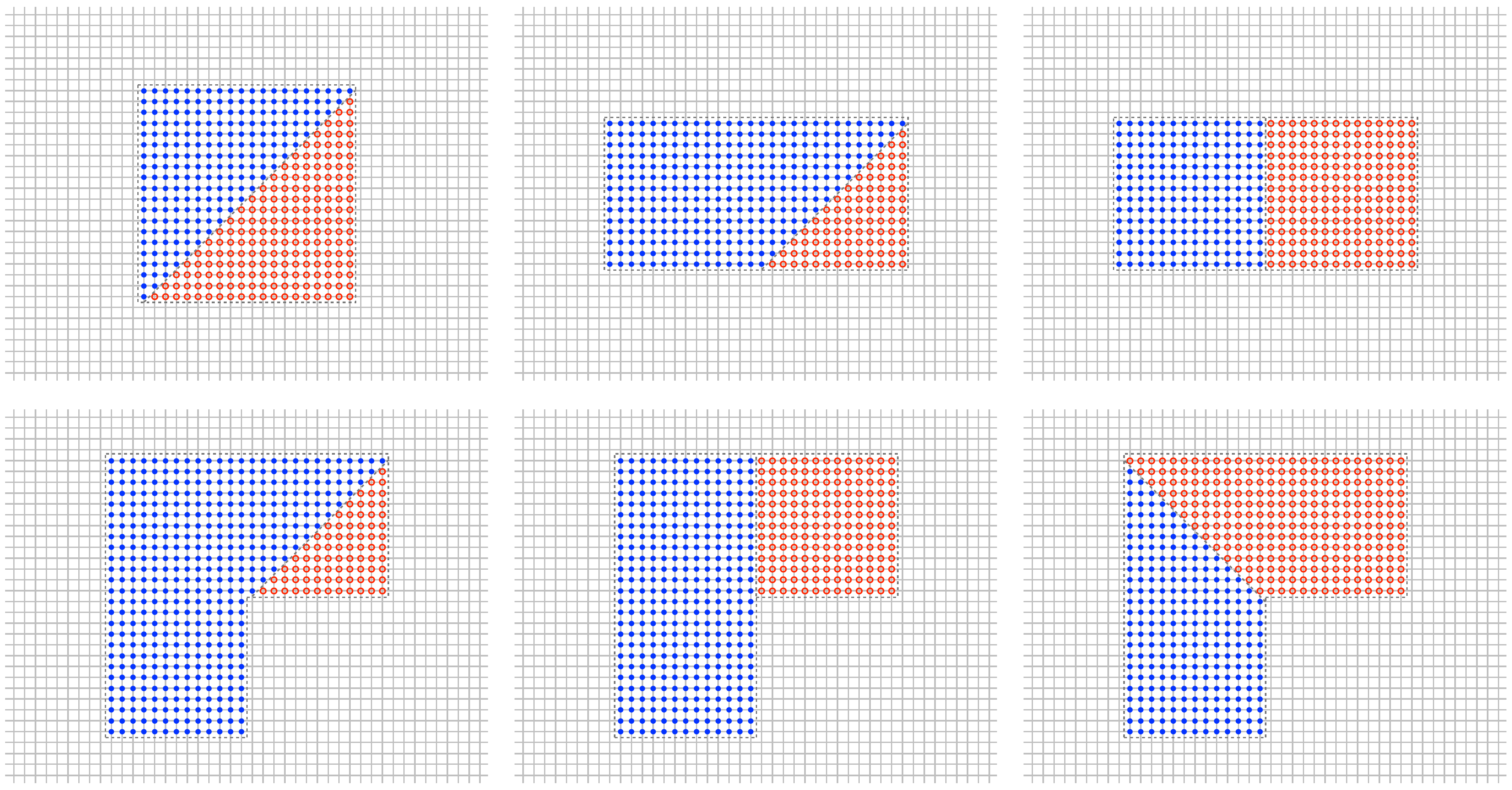}
\vspace{-.4cm}
\caption{Configurations of adjacent domains on the lattice, identified by red circles and blue dots, which have been employed to find $b(\theta^{(1)}, \theta^{(2)}) $ for some values of arguments such that $\theta^{(1)}+  \theta^{(2)} < 2\pi$ (coloured markers in Fig.\,\ref{fig:cornerNegNew}).
}
\label{fig:configs2}
\end{figure}

Considering the scaling limit for the configurations of adjacent domains $A_1$ and $A_2$ in Figs.\,\ref{fig:configs1}, \ref{fig:configs2} and \ref{fig:configs3} where $\partial A_1$ and $\partial A_2$ contain vertices, 
the expansion of the logarithmic negativity contains a subleading logarithmic term after the leading area law term.
By analogy with the case of the entanglement entropy, it is reasonable to expect that the coefficient of the logarithmic term is obtained by summing the contributions of the vertices occurring in $\partial A_1$ and $\partial A_2$.
Extracting the coefficient of such subleading logarithmic term from the lattice numerical data is a delicate task.

Let us first discuss the method employed to get the numerical values presented in this section from the formulas discussed in \S\ref{sec:hc}.
A first rough approach could consist in fitting the numerical data by a linear term, a logarithmic term and a constant one.
Nevertheless, since the logarithmic contribution is tiny with respect to the linear one,  its estimation can be spoiled by the occurrence of subleading lattice effects.
In order to take them into account, we have inserted in our fitting analysis also some standard power law corrections, i.e. we have fitted our data with a function of the form
$c_1\ell + c_\text{log}\log(\ell) + c_0 + c_{-1}\ell^{-1}+ \dots + c_{-k_\text{\tiny max}}\ell^{-k_\text{\tiny max}}$ \cite{ch-06-corners,ch-review},
where $\ell$ is a characteristic length of $P_{\text{\tiny shared}}$.
In particular, considering the domains identified by red circles in Figs.\,\ref{fig:configs1}, \ref{fig:configs2} and \ref{fig:configs3}, in each configuration $\ell$ corresponds to the edge for squares, to the shortest edge for the triangles, to the radius for the disks and to the height for the trapezoids.
For each data set, in the plots  we show the results of various fits performed in different ranges of $\ell$, where
each range is specified by the starting value $\ell_\text{\tiny start}$ and by the ending value $\ell_\text{\tiny end}$.
We fix some maximum exponent $k_\text{\tiny max}$ and some $\ell_\text{\tiny start}$ by removing some initial points, which are strongly affected by lattice effects and therefore would need more corrections.
Then, we plot the logarithmic coefficient obtained from different fits as a function of $\ell_\text{\tiny end}$.
The parameters $k_\text{\tiny max}$, $\ell_\text{\tiny start}$ and the maximum value of $\ell_\text{\tiny end}$ are chosen in order to get a stable result from the fits in the whole range of $\ell_\text{\tiny end}$.
The logarithmic coefficient is estimated as the average of the fitted values within such range of $\ell_\text{\tiny end}$.
The error introduced through this procedure is estimated  by taking the maximum deviation of the data from the average within this range of stability.

An important benchmark employed to test our numerical analysis is the mutual information of adjacent domains.
In particular, we have considered the coefficient of the logarithmic term in the mutual information for the configurations in the 
top left, bottom middle and bottom right panels of Fig.\,\ref{fig:configs1}, recovering the values of the corner function $\tilde{b}(\theta)$ for $\theta \in \{\pi/4, \pi/2 , 3\pi/4\}$ available in the literature \cite{ch-06-corners, ch-review, bueno-15}.

Let us consider the occurrence of a subleading logarithmic term due to corners in the logarithmic negativity of some configurations of adjacent domains. 
We first observe that the angles of $A_1$ and $A_2$ contributing to the logarithmic term in $\mathcal{E}$ are such that at least one of their sides belongs to $\partial A_1\cap \partial A_2$, namely only the angles whose vertices lie on the curve $\partial A_1\cap \partial A_2$ provide a non trivial logarithmic contribution. Also the endpoints of $\partial A_1\cap \partial A_2$ have to be included among such vertices. 
This is expected from the guiding principle that the logarithmic negativity measures the entanglement between $A_1$ and $A_2$.

A way to check numerically this observation is to consider e.g. the coefficients of the logarithmic terms in $\mathcal{E}$ for the configurations in the top panels of Fig.\,\ref{fig:configs1}. 
These configurations contain only three possible different contributions corresponding to this kind of vertices.
Thus, one can easily solve the resulting linear system of three equations finding that the contribution coming from the four vertices which do not belong to $\partial A_1\cap \partial A_2$ is much smaller than the other ones (by a factor of about $1/100$).
As further check that only the vertices lying on $\partial A_1\cap \partial A_2$  (including its endpoints) contribute to the logarithmic term of $\mathcal{E}$, we have constructed another configuration of adjacent domains as follows:
starting from a configuration like the one in the top right panel of Fig.\,\ref{fig:configs1} with $\ell_{\textrm{\tiny in}} / \ell_{\textrm{\tiny out}} = 1/3$ and dividing the domain corresponding to the blue dots along the diagonal with negative slope of $A_1\cap A_2$, we have removed the upper triangle. 
In the resulting configuration $A_1\cup A_2$ is a triangle and the subregion identified by the red circles is a square. 
Comparing the coefficients of the logarithmic term for this configuration and the one for the configuration in the top right panel of Fig.\,\ref{fig:configs1} with $\ell_{\textrm{\tiny in}} / \ell_{\textrm{\tiny out}} = 1/3$, we have found the same number within  numerical errors.

\begin{figure}[t!]
\vspace{.2cm}
\hspace{1.5cm}
\includegraphics[width=.8\textwidth]{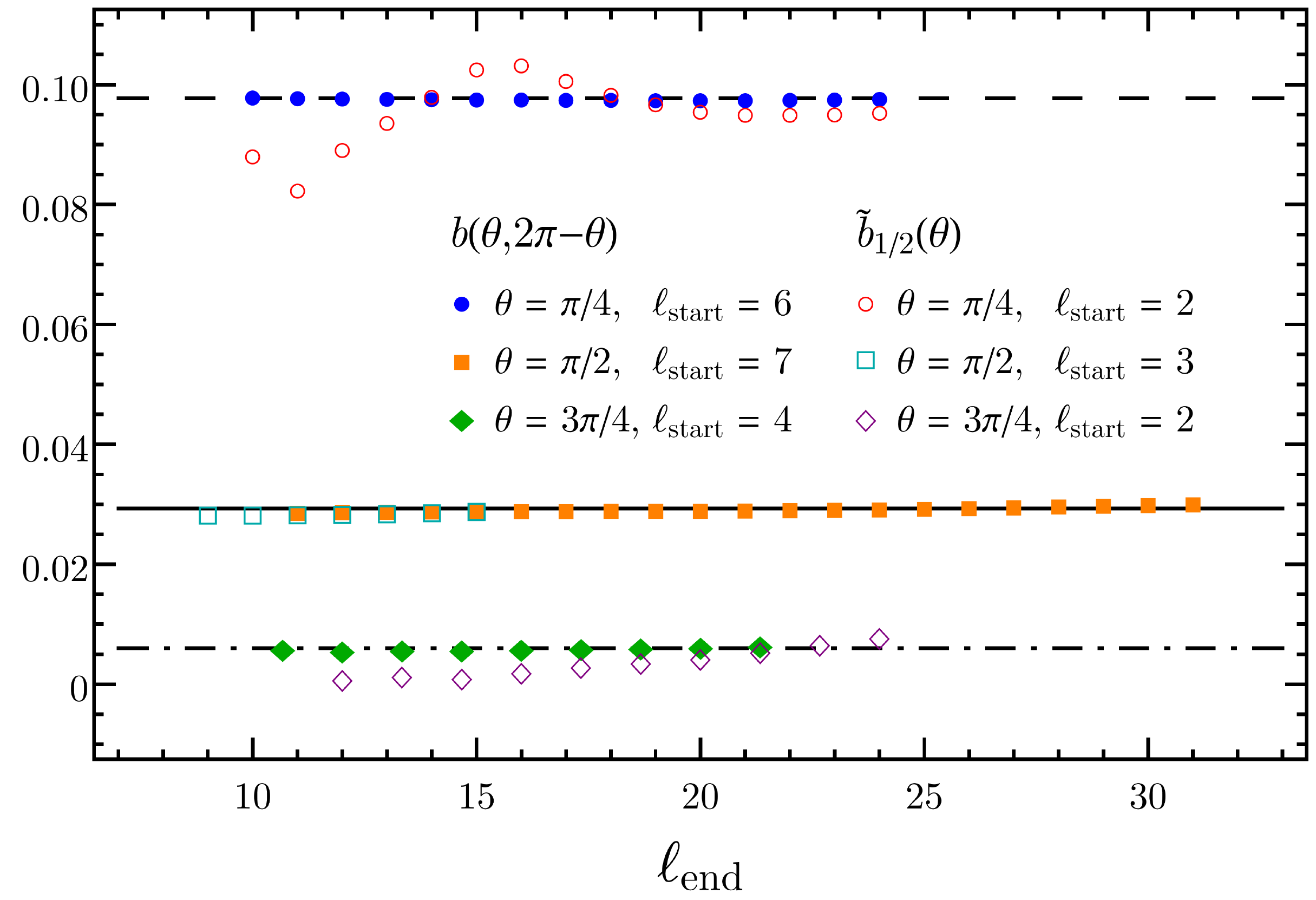}
\vspace{.0cm}
\caption{Stability analysis of the fitted values of the corner functions $b(\theta,2\pi -\theta)$  and $\tilde{b}_{1/2}(\theta)$  for $\theta \in \{\pi/4, \pi/2 , 3\pi/4\}$, as explained in \S\ref{sec:corner logneg}.
The configurations employed here are shown in the top left, bottom middle and bottom right panels of Fig.\,\ref{fig:configs1}.
The horizontal lines (with various dashing) correspond to the estimates obtained as explained in \S\ref{sec:corner logneg}. The numerical values are $b(\pi/4,7\pi/4)=0.0977(3)$, $b(\pi/2,3\pi/2)=0.029(1)$ and $b(3\pi/4,5\pi/4)=0.0060(5)$.
}
\label{fig:cornerNegOld}
\end{figure}

Thus, the logarithmic negativity of adjacent domains whose boundaries share a curve containing some vertices, where its endpoints are counted among them, is given by 
\be
\label{area law neg corners}
\mathcal{E} \,=\, a\, P_{\textrm{\tiny shared}} 
- 
\bigg(
\hspace{-.03cm}
\sum_{{\textrm{vertices of}} \atop {\textrm{$\partial A_1 \cap \partial A_2$}}} 
\hspace{-.2cm}
b(\theta_i^{(1)},  \theta_i^{(2)})
\bigg)
\log P_{\textrm{\tiny shared}}  + \dots\,,
\ee
being $\theta_i^{(k)}$ the angle corresponding to the $i$-th vertex of $\partial A_1 \cap \partial A_2$ which belongs to $A_k$.
In (\ref{area law neg corners}) we have assumed that either two or three curves meet at every vertex of $\partial A_1 \cap \partial A_2$, i.e. every vertex corresponds either to a bipartition ($\theta_i^{(1)} + \theta_i^{(2)} =2\pi$) or to a tripartition ($\theta_i^{(1)} + \theta_i^{(2)} < 2\pi$)  respectively of the angle of  $2\pi$.
This assumption, which has been done also in (\ref{area law MI corners}), is verified for all the configurations in  Figs.\,\ref{fig:configs1}, \ref{fig:configs2} and \ref{fig:configs3}.

From our numerical analysis, we find that the above considerations apply also for the quantity $\mathcal{E}_n $ defined in (\ref{En def}). Thus we have
\be
\label{area law renyi neg corners}
\mathcal{E}_n \,=\, a_n\, P_{\textrm{\tiny shared}} 
- 
\bigg(
\hspace{-.03cm}
\sum_{{\textrm{vertices of}} \atop {\textrm{$\partial A_1 \cap \partial A_2$}}} 
\hspace{-.2cm}
b_n(\theta_i^{(1)},  \theta_i^{(2)})
\bigg)
\log P_{\textrm{\tiny shared}}  + \dots\,,
\ee
where the coefficient $a_n$ has been already discussed in \S\ref{sec: area neg renyi} and the corner function $b_n(\theta_i^{(1)},  \theta_i^{(2)})$ is related to the one occurring in the logarithmic term of (\ref{area law neg corners}) through the replica limit (\ref{replica limit E_n}), namely $\lim_{n_e \to 1} b_{n_e}(\theta_i^{(1)},  \theta_i^{(2)}) = b(\theta_i^{(1)},  \theta_i^{(2)})$.

Among the vertices belonging to the curve $\partial A_1 \cap \partial A_2$ which contribute to the logarithmic term in (\ref{area law neg corners}) and (\ref{area law renyi neg corners}), let us consider first the ones corresponding to pairs of explementary angles, i.e.\ the ones such that $\theta_i^{(1)}+  \theta_i^{(2)} = 2\pi$.
This kind of vertices occurs in all the panels of Fig.\,\ref{fig:configs1} except for the bottom left one (in particular, for the configurations in the top left, bottom middle and bottom right panels only this kind of vertices occurs), while it does not occur at all in the configurations of Fig.\,\ref{fig:configs2}.

For these vertices we can make an observation similar to the one that leads to (\ref{area law coeff a_n}).
Indeed, because of the local nature of the function $b_n(\theta_i, 2\pi-\theta_i)$, it is reasonable to assume that these vertices provide the same contribution given in the case of a bipartition of the ground state, when (\ref{identity pure states}) holds. 
This observation leads us to propose the following relation between $b_n(\theta, 2\pi-\theta)$ and the corner function $\tilde{b}_n(\theta) $ entering in the R\'enyi entropies
\be
\label{corner function pure}
b_n(\theta, 2\pi-\theta) =  
  \Bigg\{\begin{array}{ll}
\big(1-n_o\big)\, \tilde{b}_{n_o}(\theta)  & \hspace{.5cm}  \textrm{odd $n=n_o$}\,,
\\
\rule{0pt}{.6cm}
2 \big(1-n_e/2\big)\,\tilde{b}_{n_e/2}(\theta)  &  \hspace{.5cm}  \textrm{even $n=n_e$}\,.
\end{array}
\ee
By employing the replica limit (\ref{replica limit E_n}), the relation (\ref{corner function pure}) allows to conclude that the corner function in the logarithmic negativity for this kind of vertices is equal to the corner function in the R\'enyi entropy of order $1/2$, namely
\be
\label{conj corner neg 1/2}
b(\theta, 2\pi - \theta) =  \tilde{b}_{1/2}(\theta)\,.
\ee
Numerical checks of the relation (\ref{conj corner neg 1/2}) for some values of $\theta$ are shown in Fig.\,\ref{fig:cornerNegOld}.
The values of $b(\theta, 2\pi - \theta) $ and $\tilde{b}_{1/2}(\theta)$ for $\theta \in \{\pi/4, \pi/2, 3\pi/4\}$ have been found by evaluating $\mathcal{E}$ and $I^{(1/2)}_{A_1, A_2}$ for the configurations shown in the top middle, bottom middle and bottom right panels of Fig.\,\ref{fig:configs1}, where the curve $\partial A_1 \cap \partial A_2$ contains only the kind of vertices that we are considering. 
The numerical values obtained for $b(\theta, 2\pi - \theta) $ for the above opening angles are: $b(\pi/4,7\pi/4)=0.0977(3)$, $b(\pi/2,3\pi/2)=0.029(1)$ and $b(3\pi/4,5\pi/4)=0.0060(5)$.
The corresponding numerical values obtained for $\tilde{b}_{1/2}(\theta)$ are less stable than the ones for $b(\theta, 2\pi - \theta) $. 
Nevertheless, they provide a reasonable check of (\ref{conj corner neg 1/2}).

\begin{figure}[t!]
\vspace{.3cm}
\hspace{.83cm}
\includegraphics[width=.8\textwidth]{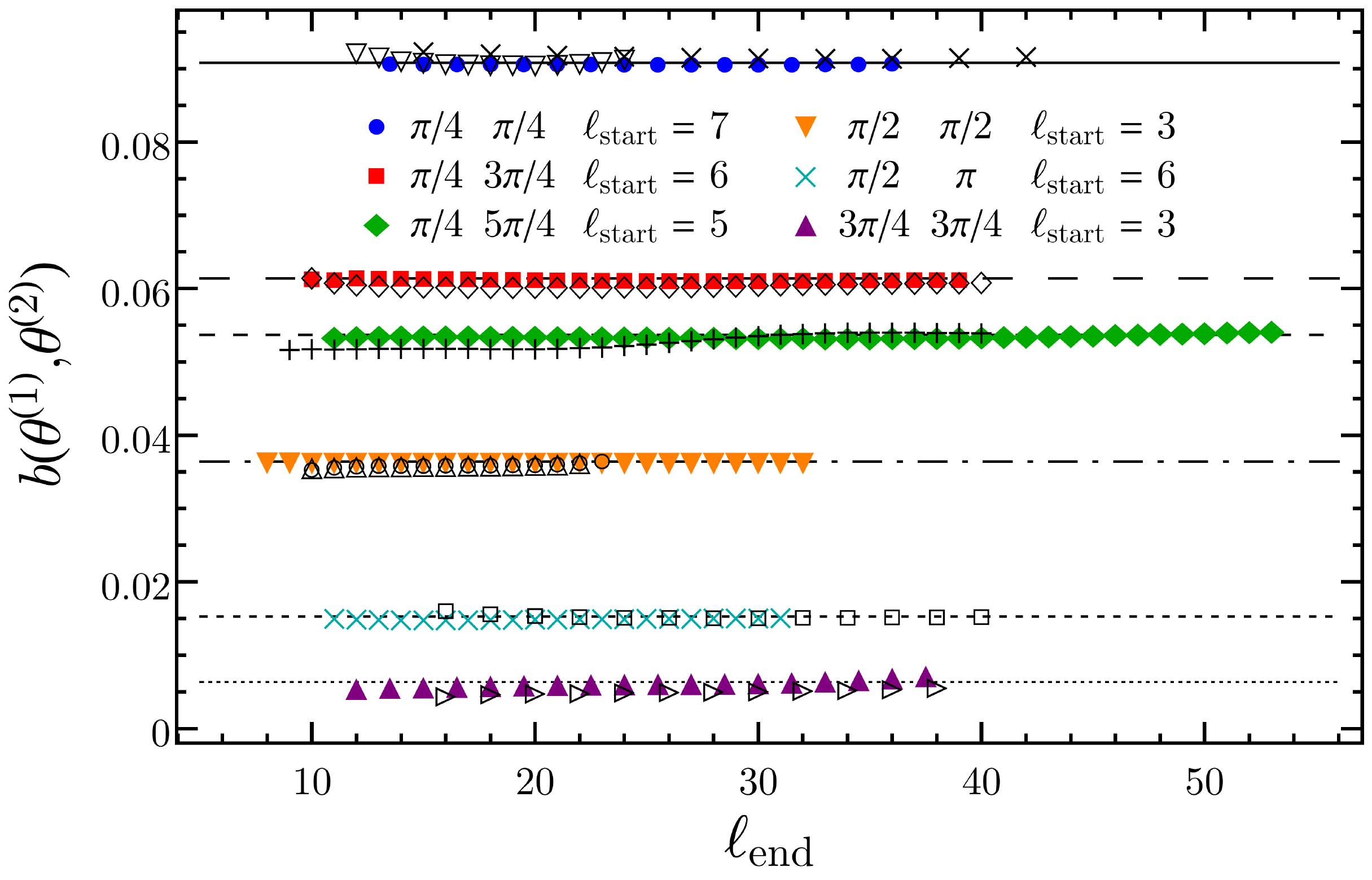}
\vspace{-.1cm}
\caption{Stability analysis of the fitted values of the corner function $b(\theta^{(1)},\theta^{(2)})$ for some pair of angles such that $\theta^{(1)} + \theta^{(2)} < 2\pi$. 
The data corresponding to coloured markers have been obtained by employing the configurations in Fig.\,\ref{fig:configs2}, 
while the data corresponding to black markers have been found by using the configurations in Fig.\,\ref{fig:configs3} and the ones in the top middle and 
top right panels of Fig.\,\ref{fig:configs1}.
The horizontal lines (with various dashing) correspond to the estimates obtained as explained in \S\ref{sec:corner logneg}. The numerical values are reported in (\ref{corner tri-vertex values}).
}
\label{fig:cornerNegNew}
\end{figure}

An analytic expression for the function $\tilde{b}_{1/2}(\theta)$, which can be found by performing the analytic continuation $n \to 1/2$ of the formula for $\tilde{b}_{n}(\theta)$ obtained in \cite{ch-06-corners}, is not available.
Considering the expansion of the corner function $\tilde{b}_n(\theta) = \sigma_n (\pi- \theta)^2 + \dots$ as $\theta \to \pi$, where the dots denote subleading contributions, in \cite{bueno-15} it has been found that the leading term provides a lower bound, namely $\tilde b_n(\theta)\geqslant \tilde b^\text{\tiny l.b.}_n(\theta)$, where $\tilde b^\text{\tiny l.b.}_n(\theta)=\sigma_n(\theta-\pi)^2$.
The coefficient $\sigma_n$ has been computed for simple models like the Dirac fermion and the complex scalar for integer $n$ (see e.g. Table 2 of \cite{bmk-jhep}) and a duality between the free bosonic and fermionic contributions allows to get also $\sigma_{1/n}$ \cite{bmk-jhep}.
In particular, for the real free boson one gets $\sigma_{1/2}=\tfrac{1}{32\pi}$ (see Table 5 of \cite{bmk-jhep}), which leads to the following lower bounds for $\tilde b_{1/2}(\theta)$ for the opening angles that we analysed: 
$\tilde b^\text{\tiny l.b.}_n(\pi/4)=0.0552$, $\tilde b^\text{\tiny l.b.}_n(\pi/2)=0.0245$ and $\tilde b^\text{\tiny l.b.}_n(3\pi/4)=0.00614$.
Our numerical values for $b(\theta,2\pi-\theta)$ are above these limiting values. 
The bound becomes stronger and stronger as the angle $\theta$ approaches $\pi$, as expected from the fact that the corner function decreases monotonically to zero.
For $\theta=3\pi/4$, our result satisfies the bound only once the estimated error is taken into account.

\begin{figure}[t!]
\vspace{.1cm}
\hspace{-.7cm}
\includegraphics[width=1.08\textwidth]{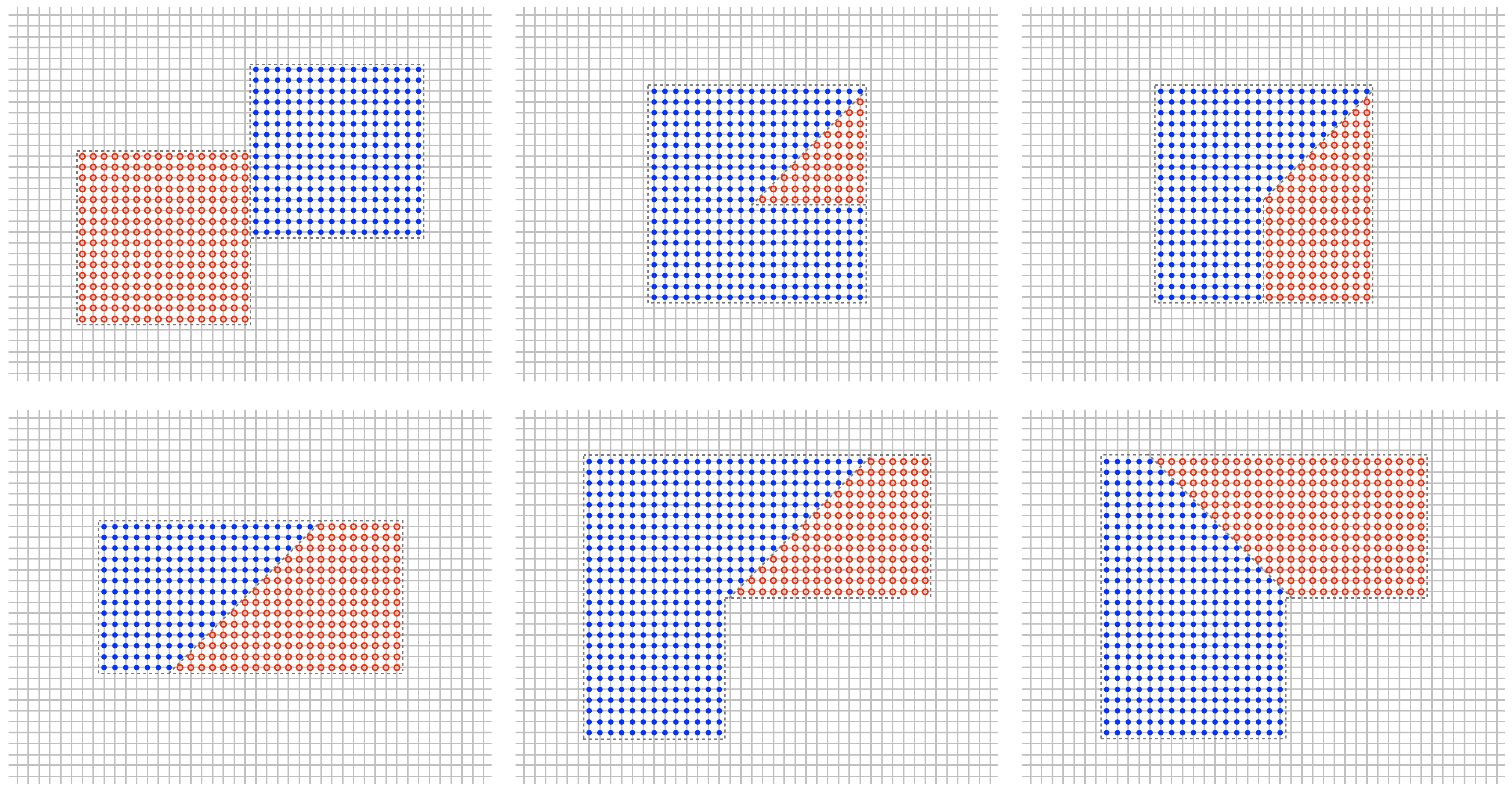}
\vspace{-.4cm}
\caption{Configurations of adjacent domains on the lattice, identified by red circles and blue dots, which have been employed as crosschecks for the values of $b(\theta^{(1)}, \theta^{(2)}) $ given in Fig.\,\ref{fig:cornerNegOld} and Fig.\,\ref{fig:cornerNegNew}. 
}
\label{fig:configs3}
\end{figure}

Let us consider the vertices corresponding to a tripartition of the angle $2\pi$, for which $\theta_i^{(1)}+  \theta_i^{(2)} < 2\pi$ and which are the endpoints of the curve $\partial A_1 \cap \partial A_2$ in the class of vertices that we are considering. 
For this kind of vertices the corner function $b_n(\theta_i^{(1)}, \theta_i^{(2)})$ depends on two independent variables. 
We have obtained its numerical values for some pairs of angles by employing the configurations in Fig.\,\ref{fig:configs2}.
The results are shown in Fig.\,\ref{fig:cornerNegNew} (coloured solid markers) and they are given by
\be
\label{corner tri-vertex values}
\begin{array}{lll}
b(\pi/4,\pi/4) = 0.0908(1) \hspace{.5cm}
 & b(\pi/4,3\pi/4) =  0.0614(3) \hspace{.5cm}
 &  b(\pi/4,5\pi/4) = 0.0536(1)
 \\
b(\pi/2,\pi/2) = 0.0364(1)  \hspace{.5cm}
 & b(\pi/2, \pi) =  0.0152(2) \hspace{.5cm}
 &  b(3\pi/4,3\pi/4) = 0.0068(2)\,,
\end{array}
\ee
where the parenthesis denote the uncertainty on the last digit. 
In Fig.\,\ref{fig:cornerNegNew} the black markers correspond to the values obtained by employing all the configurations of Fig.\,\ref{fig:configs3} and the ones in the top middle and top right panels of Fig.\,\ref{fig:configs1}.

Let us remark that the coefficient of the logarithmic term in the logarithmic negativity of the configuration in the top left panel of Fig.\,\ref{fig:configs3} is non zero, while it is vanishing in the mutual information of the same configuration, as already pointed out in \S\ref{sec:corners ent}.
Moreover, such coefficient for the configuration shown in the the bottom left panel of Fig.\,\ref{fig:configs1} turns out to be zero within our numerical errors, both for the negativity and the mutual information, as expected from the fact that corners do not occur in the continuum limit.

In principle, our numerical analysis allows  to find also $b_n(\theta_i^{(1)}, \theta_i^{(2)})$.
Nevertheless,  in order to check the relation (\ref{corner function pure}) we need to know the unusual corrections \cite{cc-10} to the scaling in $2+1$ dimensions in order to perform a precise fitting analysis (see e.g. \cite{melko-15}).

\subsection{Comments on the continuum limit}

The continuum limit of the lattice model (\ref{ham}) considered throughout this manuscript is described by the massless scalar field in $2+1$ dimensions, which is a CFT. 
Denoting by $\varepsilon$ the UV cutoff that must be introduced to regularise the model, the logarithmic negativity of adjacent domains diverges when $\varepsilon \to 0$.

The numerical results on the lattice discussed above tell us that the expansion of the logarithmic negativity as $\varepsilon \to 0$ reads
\be
\label{area law neg corners qft}
\mathcal{E} \,=\, \alpha\, \frac{P_{\textrm{\tiny shared}}}{\varepsilon }
- 
\bigg(
\hspace{-.03cm}
\sum_{{\textrm{vertices of}} \atop {\textrm{$\partial A_1 \cap \partial A_2$}}} 
\hspace{-.2cm}
b(\theta_i^{(1)},  \theta_i^{(2)})
\bigg)
\log (P_{\textrm{\tiny shared}} / \varepsilon ) + \dots\,,
\ee
where the coefficient $\alpha$ in front of the area law term is non universal.
A similar expression can be written for $\mathcal{E}_n$ in (\ref{En def}), namely
\be
\label{area law renyi neg corners qft}
\mathcal{E}_n \,=\, \alpha_n\, \frac{P_{\textrm{\tiny shared}}}{\varepsilon }
- 
\bigg(
\hspace{-.03cm}
\sum_{{\textrm{vertices of}} \atop {\textrm{$\partial A_1 \cap \partial A_2$}}} 
\hspace{-.2cm}
b_n(\theta_i^{(1)},  \theta_i^{(2)})
\bigg)
\log (P_{\textrm{\tiny shared}} / \varepsilon ) + \dots\,,
\ee
where $\alpha_n$ is non universal as well.
Instead, the corner functions $b(\theta_i^{(1)},  \theta_i^{(2)})$ and $b_n(\theta_i^{(1)},  \theta_i^{(2)})$ are independent of the UV details of the regularisation and therefore they are the most important quantities to study.
In (\ref{area law neg corners qft}) and (\ref{area law renyi neg corners qft}), like for their lattice versions (\ref{area law neg corners}) and (\ref{area law renyi neg corners}), we have assumed that the vertices in $\partial A_1$ and $\partial A_2$  correspond either to a bipartition or to a tripartition of the angle of $2\pi$.

The divergent terms in the $\varepsilon \to 0$ expansion of $\mathcal{E}$ and $\mathcal{E}_n$ are determined by local effects close to the curve $\partial A_1 \cap \partial A_2$, consistently with the intuition that the entanglement between $A_1$ and $A_2$ comes from the degrees of freedom living close to their shared boundary.
This leads to relate the coefficients $\alpha$ and $\alpha_n$ in (\ref{area law neg corners qft}) and (\ref{area law renyi neg corners qft}) to the area law coefficients $\tilde\alpha$ and $\tilde\alpha_n$ of $S_A$ and $S_A^{(n)}$ like in (\ref{area law coeff a_n}) and (\ref{area coeff neg renyi 1/2}).
Notice that, whenever for the $i$-th vertex of $\partial A_1 \cap \partial A_2$ we have $\theta_i^{(1)} + \theta_i^{(2)}=2\pi$, the relations (\ref{corner function pure}) and (\ref{conj corner neg 1/2}) for the corner functions $b(\theta,  2\pi -\theta)$ and $b_n(\theta,  2\pi -\theta)$ hold also in the continuum limit. 
Instead, whenever the vertices correspond to partitions of the angle of $2\pi$ in three (i.e. $\theta_i^{(1)} + \theta_i^{(2)} < 2\pi$) or higher number of components, we expect that the corner functions occurring in  $\mathcal{E}$ or $\mathcal{E}_n$ contain new information with respect to the corner functions  entering in $S_A$ or $S_A^{(n)}$.

\begin{figure}[t!]
\vspace{.2cm}
\hspace{-1.1cm}
\includegraphics[width=1.15\textwidth]{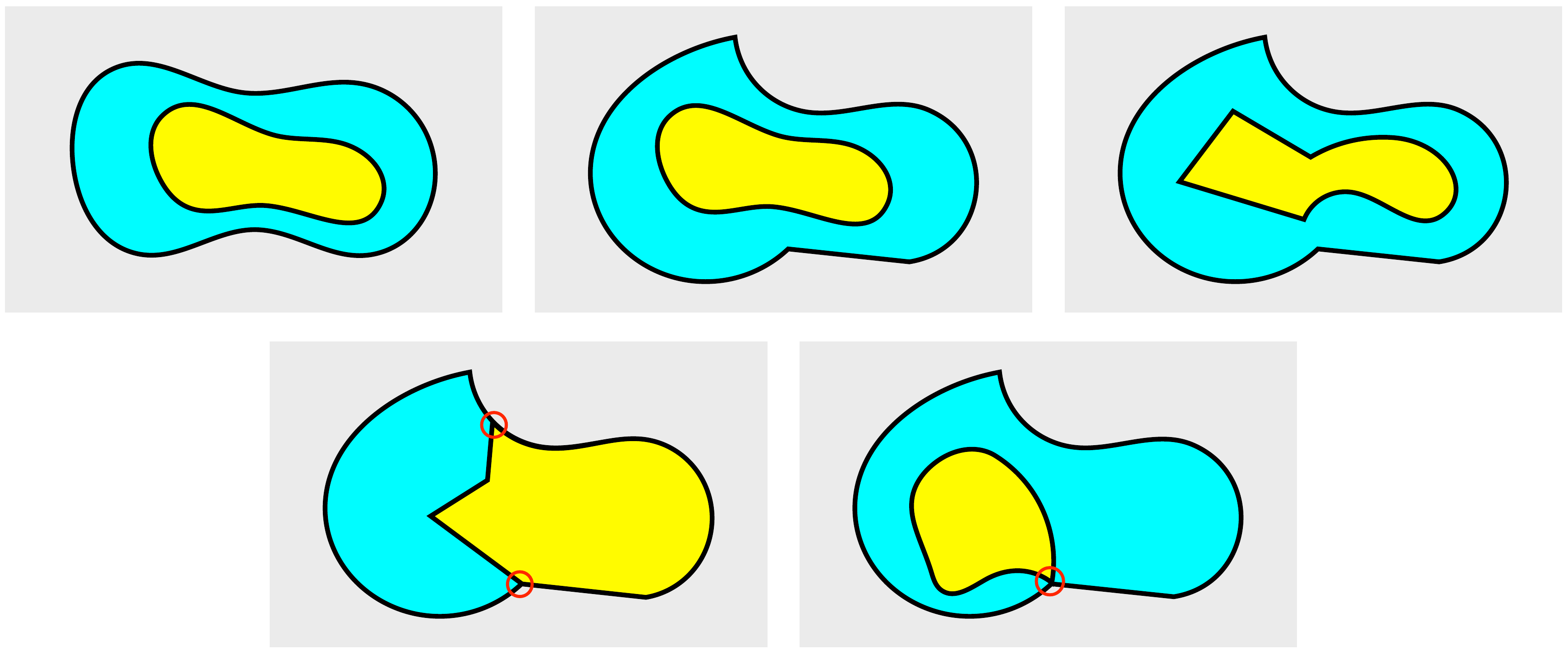}
\vspace{-.4cm}
\caption{Configurations of adjacent domains $A_1$ and $A_2$ in the plane (the yellow and cyan regions). 
The grey part corresponds to the region $B$ which has been traced out. 
The quantities $\mathcal{E}$ and $\mathcal{E}_n$ for the configurations in the top left and top middle panels do not contain a logarithmic term, while for the remaining ones such term is non vanishing. 
The vertices corresponding to a partition of the angle of $2\pi$ in three or four angles are highlighted
with red circles. 
}
\label{fig:configsqft}
\end{figure}

It would be very interesting to find an analytic expression for the corner function $b(\theta, \gamma)$ for any pair of angles $\theta$ and $\gamma$, where we can assume $\theta \leqslant \gamma$.
By considering only the few pairs of angles that we have studied on the lattice, we can make some observations about the behaviour of this corner function. 
For instance, when the two angles are equal, the function $b(\theta,\theta)$ is decreasing with $\theta\in (0,\pi]$ and $b(\pi,\pi) = 0$.
Moreover, for fixed $\theta \leqslant \pi$, the function $b(\theta,\gamma)$ is decreasing with $\gamma\in(\theta ,2\pi-\theta)$.
By comparing Fig.\,\ref{fig:cornerNegNew} with Fig.\,\ref{fig:cornerNegOld}, one notices that this is not true anymore when $\gamma$ is exactly $2\pi-\theta$.
Such behaviour is not surprising because in this limit the boundaries of $A_1$ and $A_2$ merge, $P_\text{\tiny shared}$ changes abruptly and therefore a continuous behaviour of the divergent terms in the logarithmic negativity is not expected.

In Fig.\,\ref{fig:configsqft} we show some illustrative examples of configurations of adjacent regions in the plane.
The grey region is associated to the part $B$, which has been traced out. 
Considering $\mathcal{E}$ and $\mathcal{E}_n$ between the yellow domain and the cyan domain, the expressions (\ref{area law neg corners qft}) and (\ref{area law renyi neg corners qft}) can be employed  for all the configurations in  Fig.\,\ref{fig:configsqft}  except for the one in the bottom right panel, where the vertex highlighted by the red circle corresponds to a partition of the angle of $2\pi$ in four parts.

In the configurations shown in the top left and middle panels of Fig.\,\ref{fig:configsqft} the curve $\partial A_1 \cap \partial A_2$ is smooth; therefore the logarithmic divergence does not occur in $\mathcal{E}$ and $\mathcal{E}_n$.
Instead, in the remaining configurations the curve $\partial A_1 \cap \partial A_2$ contains vertices and the subleading logarithmic divergence occurs. 
As for the configuration in the top right panel, the coefficient of the logarithmic term of $\mathcal{E}$ and $\mathcal{E}_n$ is related to the corner functions $\tilde{b}(\theta)$  and $\tilde{b}_n(\theta)$ entering in the logarithmic term of $S_A$ or $S_A^{(n)}$ respectively through (\ref{corner function pure}) and (\ref{conj corner neg 1/2}).
Thus, for the configurations in the top panels of Fig.\,\ref{fig:configsqft} and assuming that only one scale $\ell$ occurs to determine the logarithmic term, one can construct the following UV finite quantity
\be
\label{neg minus mi}
\mathcal{E}  - \frac{1}{2}\, I^{(1/2)}_{A_1, A_2}\,.
\ee
For the same configurations, also the following combinations, depending on the parity of the integer $n$, are UV finite 
\be
\label{renyi neg minus mi}
\mathcal{E}_{n_o} -  \frac{1-n_o}{2}\, I^{(n_o)}_{A_1, A_2} \,,
 \hspace{.3cm} \qquad \hspace{.3cm}
 \mathcal{E}_{n_e} -  \left(1-\frac{n_e}{2}\right) I^{(n_e/2)}_{A_1, A_2}\,.
\ee
The second expression in (\ref{renyi neg minus mi}) provides (\ref{neg minus mi}) after the analytic continuation $n_e \to 1$.
In a $2+1$ dimensional CFT, the quantities in (\ref{neg minus mi}) and (\ref{renyi neg minus mi}) should give non trivial scale invariant functions of the parameters characterising the adjacent domains. 
For example, when the adjacent domains are given by a disk of radius $R_{\textrm{\tiny in}}$  and an annulus surrounding it whose radii are $R_{\textrm{\tiny in}} < R_{\textrm{\tiny out}}$, the expression (\ref{neg minus mi}) should be a  model dependent function of the ratio $R_{\textrm{\tiny in}} / R_{\textrm{\tiny out}}$.
It would be very interesting to develop a method which allows to get an analytic expression for this function.

The configurations of adjacent domains shown in the bottom panels of Fig.\,\ref{fig:configsqft} are more interesting because the corner functions corresponding to the vertices highlighted by the red circles is not related to the corner functions occurring in the entanglement entropies. 
Thus, because of such terms, we expect that the combinations in (\ref{neg minus mi}) and (\ref{renyi neg minus mi}) diverge logarithmically for these configurations.

\section{Conclusions}
\label{sec:conclusions}

In this manuscript we have investigated the logarithmic negativity $\mathcal{E}$ and the moments of the partial transpose for adjacent domains
$A_1$ and $A_2$ in the ground state of a two dimensional harmonic square lattice with nearest neighbour spring-like interaction.
The regime of massless oscillators in the thermodynamic limit has been considered.

By exploring various configurations of adjacent domains,  we have shown that, at leading order for large domains, the logarithmic negativity and  the quantity $\mathcal{E}_n$ introduced in (\ref{En def}) satisfy an area law
in terms of the length of the shared curve $\partial A_1 \cap \partial A_2$, suggesting a relation between the coefficient of the area law term in these quantities and the coefficient of the area law term in the R\'enyi entropies. 

A subleading universal logarithmic term occurs in $\mathcal{E}$ and $\mathcal{E}_n$ whenever the shared curve contains vertices,
being its endpoints included among them.
The values of the corner function of $\mathcal{E}$ have been obtained for some pairs of angles. 
For the vertices of $\partial A_1 \cap \partial A_2$ corresponding to pairs of explementary angles, we have proposed that the corner function of $\mathcal{E}_n$ is related to the corner function entering in the R\'enyi entropies \cite{ch-06-corners}.
This relation implies that the corner function of $\mathcal{E}$ for this kind of vertices coincides with the corner function of the R\'enyi entropy of order $1/2$. 
This statement has been supported by numerical evidences shown in Fig.\,\ref{fig:cornerNegOld}.

As for the vertices of the curve $\partial A_1 \cap \partial A_2$ corresponding to a tripartition of the angle of $2\pi$, their contribution to the logarithmic term in $\mathcal{E}$ and $\mathcal{E}_n$ is characterised by a new corner function which depends on two independent angular variables. 
The numerical values of this corner function for $\mathcal{E}$ have been given in (\ref{corner tri-vertex values}) for some pair of angles (see also Fig.\,\ref{fig:cornerNegNew}).

Let us conclude with some open questions.
It would be interesting to provide further checks of (\ref{conj corner neg 1/2}).
In particular, the analytic continuation to $n=1/2$ of the corner function found by Casini and Huerta \cite{ch-06-corners} should be performed.
More importantly, a method should be found to compute analytically the corner functions $b(\theta,\gamma)$ and $b_n(\theta,\gamma)$ for the vertices partitioning the angle of $2\pi$ in three parts. 
By analogy with the results of \cite{corner-recent qft} obtained for the corner function of $S_A$, it could be interesting to study the corner function $b(\theta, \theta) $ for equal angles as $\theta \to \pi^-$.
In order to extract reliable numerical results for the logarithmic term of $\mathcal{E}_n$ from the fit of the lattice data, the unusual corrections to the scaling must be studied, extending the analysis done by Cardy and Calabrese \cite{cc-10} in $1+1$ dimensions. 
Finally, it is worth studying the corner contributions to $\mathcal{E}$  and $\mathcal{E}_n$ for other models, both on the lattice and in the continuum.

\section*{Acknowledgements}

We are grateful in particular to Pasquale Calabrese, John Cardy, Horacio Casini and Marina Huerta for useful discussions and correspondence. 
We thank Vincenzo Alba, Mario Collura, Viktor Eisler, Veronika Hubeny, Mukund Rangamani, German Sierra, William Witczak-Krempa and Zoltan Zimbor\'as for useful conversations. 
ET would like to thank the Instituto de Fisica Teorica (IFT UAM-CSIC) in Madrid for hospitality during part of this work and its support via the Centro de Excelencia Severo Ochoa Program under Grant SEV-2012-0249. 
ET is grateful to Princeton University, MIT, Brandeis University and UC Davis for the warm hospitality during the final part of this work.
AC acknowledges support from MINECO (grant MTM2014-54240-P) and Comunidad de Madrid (grant QUITEMAD+-CM, ref. S2013/ICE-2801).
AC has received funding from the European Research Council (ERC) under the European Union's Horizon 2020 research and innovation programme (grant agreement No 648913).
ET has been supported by the ERC under Starting Grant  279391 EDEQS.

\end{document}